\documentclass[twocolumn,superscriptaddress,showpacs,preprintnumbers,amsmath,amssymb,floatfix]{revtex4-2}

\usepackage{graphicx}
\usepackage{bm}
\usepackage{booktabs}
\usepackage{enumitem}
\usepackage{mathtools}
\usepackage{hyperref}
\usepackage{xcolor}
\usepackage{microtype}

\newcommand{\R}{\mathbb{R}}
\newcommand{\E}{\mathbb{E}}
\newcommand{\tr}{\operatorname{tr}}
\newcommand{\argmax}{\operatorname*{arg\,max}}

\newcommand{\bb}{\bm{b}}
\newcommand{\bx}{\bm{x}}
\newcommand{\bc}{\bm{c}}
\newcommand{\bs}{\bm{s}}
\newcommand{\by}{\bm{y}}

\begin{document}

\preprint{Preprint}

\title{Adaptive Sensing beyond Non-Adaptive Information Limits:\\
       End-to-End Co-Design of Geometry, Policy, and Inference}

\author{Arvin Keshvari}
\author{William Tuxbury}
\author{Zin Lin}
\email{zinlin@vt.edu}
\affiliation{Bradley Department of Electrical and Computer Engineering, Virginia Tech, Blacksburg, VA 24060, USA}

\date{\today}

\begin{abstract}
Inverse design has transformed vast physical parameter spaces into a substrate for emergent functionality, raising the tantalizing prospect of relocating intelligence from the digital domain into the physical world itself. Nowhere is this prospect more consequential than in sensing, where the analog-to-digital interface imposes a fundamental bottleneck: information not captured by the hardware is irrevocably lost to any downstream algorithm. Existing approaches improve information capture through either sensor hardware optimization or adaptive measurement strategies operating on fixed hardware, but rarely both in concert. A principled migration of intelligence from digital to physical demands their joint optimization: the sensing geometry must be co-designed with a policy that determines what to measure next. We formulate this co-design as joint dynamic programming (joint-DP), a unified optimization over sensor geometry and a Bellman-optimal adaptive measurement policy. The outer hardware gradient is obtained through differentiable dynamic programming with a sharp Bellman maximum. A hierarchy of relaxations extends the framework from small discrete POMDPs to freeform photonic topologies with more than $10^5$ design pixels.
\end{abstract}

\maketitle

\section{Introduction}
\label{sec:intro}

Modern deep learning has revealed that emergent intelligent behavior can be elicited from optimization over vast parameter spaces: neural networks with billions of weights, trained by gradient descent against differentiable objectives, produce capabilities no hand-designed architecture reaches.  A structurally analogous revolution has been unfolding in physics: adjoint-based inverse design optimizes over millions of geometric degrees of freedom against differentiable physical objectives to elicit non-intuitive, freeform structures with capabilities no hand-designed architecture can match~\cite{molesky2018inverse, jensen2011topology, fan2020freeform, christiansen2021fullwave, lin2018topology}. \emph{Structural complexity in physics is the analog of representational complexity in deep networks}: both are substrates through which optimization converts raw degrees of freedom into engineered capability. The consequence is that physical hardware optimization is not a peripheral engineering concern but a fundamental question about where intelligence can live. 

This question sharpens dramatically in sensing and metrology, at the analog-to-digital bottleneck where physical signals become a limited digital record.  Any information the hardware fails to capture at this bottleneck bounds everything any downstream algorithm can process.  The conventional blueprint of the last two decades pairs a generic, task-agnostic transducer front-end with powerful digital reconstruction, and so treats the physical hardware as a passive conduit and asks the digital stages to rescue whatever structure the geometry happens to impose. Inverse design already inverts this blueprint on the hardware side, routinely saturating the electromagnetic performance bounds of the underlying scattering problem~\cite{miller2016fundamental, miller2023bounds, molesky2018inverse, fan2020freeform}, and end-to-end optical co-design extends it by tying the geometry to a digital reconstruction backend, closing the loop from physics to inference in a single gradient~\cite{sitzmann2018e2e, chang2019diffractive, metzler2020deep, tseng2021metalens, kellman2019microscopy, heide2014flex, antipa2018diffusercam, arya2024endtoend, lin2021e2e, wang2018reconfigurable}.  Both traditions, however, still operate in a \emph{static} regime: the sensor is designed once, deployed once, and assumed throughout to produce a single one-shot measurement---a posture at odds with how real sensors function, with their \emph{tunable} settings that can be reconfigured on the fly in response to incoming data. The blueprint thus ignores the fundamentally \emph{sequential} nature of real-world sensing. What remains missing is the policy dimension: the recognition that the geometry must be co-designed with a \textit{sequential adaptive measurement policy} that reacts to intermediate data. This paper supplies this missing dimension.

A mature literature on adaptive sensing and sequential Bayesian experimental design constructs formally-optimal multi-step measurement policies for partially-observable Markov decision processes (POMDPs)~\cite{kaelbling1998pomdp, smallwood1973sondik, sondik1971optimal, kreucher2005kastella, hero2011survey, castanon2008stochastic, chen2015adaptive, krause2005near, lindley1956measure, huan2016sequential, shen2023bayesian, alexanderian2021optimal, chaloner1995bayesian, zhu2026sequentialbayesianexperimentaldesign}, but treats the physical hardware as immutable.  The hardware half (inverse design) and the policy half (POMDP theory) are thus designed separately, each community optimizing its own half against the other's default---a situation where the joint optimum is left unrealized by either side alone.  Several near-relatives live in the intersection without closing it: cognitive radar adapts the waveform on a fixed antenna~\cite{bell2015cognitive, haykin2006cognitive, greco2018cognitive, bouhou2025cognitive}, Kitaev-style quantum phase estimation adapts the pulse sequence on a fixed circuit~\cite{giovannetti2006quantum, braunstein1994sld, paris2009quantum, caves1981quantum, danilin2024scqubit, kitaev1995qpe, braverman2019slowing}, and programmable-metasurface imaging adapts the reconstruction on a fixed scattering medium~\cite{yurduseven2017printed, imani2020review}, and variational quantum sensing parametrizes the probe through a trainable neural ansatz rather than a geometric topology~\cite{pirandola2018advances, degen2017sensing}.  None of these treats continuous physical geometry and a multi-step Bellman-optimal measurement policy as co-equal design variables in a single optimization. We propose \emph{joint dynamic programming} (joint-DP), a bi-level nested co-optimization of the inner adaptive policy $\pi$ and the outer static physical design $\bc$ via bias-free gradient descent.  To our knowledge, this is the first formulation in which $\bc$ and $\pi$ are co-designed in a single end-to-end pipeline.

The ingredients of the framework are individually well-established: exact Bellman DP for POMDPs~\cite{bellman1957dynamic, sondik1971optimal, smallwood1973sondik}; end-to-end optical co-design with reconstruction backends~\cite{kellman2019physics, muthumbi2019learned, bostan2018deep, kellman2020datadriven}; differentiable dynamic programming~\cite{mensch2018differentiable, amos2018mpc, domke2012generic, blondel2022efficient}; the envelope/Danskin theorem~\cite{milgrom2002envelope, danskin1966theory, benveniste1979differentiability}; and Expected Value of Perfect Information~\cite{howard1966decision, raiffa1961applied}.  The joint optimization of continuous physical hardware with a multi-step Bellman-optimal adaptive policy is not standard in any of these; each ingredient is classical, and the novelty is their synthesis.  Our framework also complements the program on fundamental optimality limits to light--matter interaction~\cite{miller2016fundamental, miller2023bounds, chao2022maximum, molesky2020tmatrix, molesky2020hierarchical, molesky2020bounds, kuang2020maximal, gustafsson2020upper, schab2020bounds, molesky2019global, venkataram2020fundamental, jelinek2021upper}: those bounds certify whether a target is achievable by \emph{any} one-shot parametric design, but do not account for the gains available from multiple adaptive measurements. Our approach carries the guiding ambition of design optimality into the realm of sequential decision making. 

The remainder of the paper develops this framework and demonstrates it on three case studies spanning different physical domains and relaxation levels.  In Sec.~\ref{sec:framework} we define the joint-$(\bc, \pi)$ optimization and develop a family of scalar rewards compatible with the Bellman recursion.  In Sec.~\ref{sec:envelope} we show that the outer hardware gradient can be computed exactly and without bias via the envelope theorem, keeping the sharp Bellman maximum throughout.  In Sec.~\ref{sec:hierarchy} we describe a relaxation cascade that carries the framework from exact DP on small POMDPs through certainty-equivalent and myopic heuristics to a Bayesian-Fisher-maximization inner loop compatible with $10^5$-dimensional hardware optimization.  Sections~\ref{sec:radar}--\ref{sec:photonic} present the case studies: on a radar beam-search POMDP (Sec.~\ref{sec:radar}), classical information-bound-guided geometry selection loses $2.8\times$ in attainable adaptive value; on a superconducting-qubit flux sensor (Sec.~\ref{sec:scqubit}), joint-DP reduces deployed mean-squared error by $202\times$ at $z=+157\sigma$ over the joint Bayesian Cram\'er--Rao baseline, co-optimized with a five-dimensional hardware design; and on a $90{,}000$-pixel freeform photonic metasensor (Sec.~\ref{sec:photonic}), co-design with an inner policy loop maximizing the Bayesian Fisher-information reduces deployed MSE by $123\times$. We discuss limitations, generalizations and future directions in Sec.~\ref{sec:discussion}.

\section{Joint-DP framework}
\label{sec:framework}

This section defines the sensor model, the joint hardware-policy optimization, and the family of scalar rewards that make the Bellman recursion well-defined.  It then develops a value decomposition that shows why joint co-design is necessary and why the Expected Value of Perfect Information does not provide a shortcut.

\subsection{Sensor model}

A \emph{sensor} is a tuple $(\bc, \pi)$ where $\bc \in \mathcal{C} \subseteq \R^{d_c}$ is a continuous physical geometry fixed at manufacturing time (e.g., the element pattern of a phased array, the circuit parameters of a SQUID loop, the permittivity map of a metaphotonic structure) and $\pi: \mathcal{B} \to \mathcal{S}$ is a measurement policy mapping the current belief $\bb \in \mathcal{B}$ to an action $\bs \in \mathcal{S}$ (e.g., a beam pointing direction, a Ramsey delay time, phases of the probe lasers).  Here the belief $\bb_k = p(\bx \mid \bs_{1:k-1}, \by_{1:k-1};\, \bc)$ is the posterior distribution over the unknown state $\bx \in \mathcal{X}$ given all measurements collected through step $k{-}1$; at $k=1$ the belief is the prior, $\bb_1 = p_0(\bx)$.  At each step $k \in \{1, \ldots, K\}$ the policy selects an action $\bs_k = \pi(\bb_k)$, producing a measurement
\begin{equation}
  \by_k \;\sim\; p(\by_k \mid \bx, \bs_k;\, \bc),
  \label{eq:likelihood}
\end{equation}
where $\bc$ enters the likelihood through its physical forward model.  The belief is then updated via Bayes' rule,
\begin{equation}
  T(\bb, \bs, \by;\, \bc)(\bx) \;=\; \frac{p(\by \mid \bx, \bs;\, \bc)\, \bb(\bx)}{\int p(\by \mid \bx', \bs;\, \bc)\, \bb(\bx')\, d\bx'},
  \label{eq:bayes_update}
\end{equation}
giving the posterior $\bb_{k+1} = T(\bb_k, \bs_k, \by_k;\, \bc)$.  We write $K$ for the horizon, $p_0(\bx)$ for the prior, and $\pi^\star(\bc)$ for the Bellman-optimal policy at fixed $\bc$.

\subsection{Joint optimization statement}

The joint co-design objective is
\begin{equation}
  V(\bc, \pi) \;=\; \E_{\bx \sim p_0}\, \E_{\by_{1:K} \mid \bx, \pi, \bc}\!\bigl[\, \Phi(\bx, \bs_{1:K}^{\pi}, \bc) \,\bigr],
  \label{eq:V_def}
\end{equation}
where $\Phi: \mathcal{X} \times \mathcal{S}^K \times \mathcal{C} \to \R$ is a user-specified pointwise reward.  The joint optimization solves
\begin{equation}
  (\bc^\star, \pi^\star) \;=\; \argmax_{\bc \in \mathcal{C},\; \pi \in \Pi}\; V(\bc, \pi).
  \label{eq:joint_opt}
\end{equation}
Two degenerate cases recover classical programs: $\Pi = \{\text{constant policies}\}$ recovers end-to-end learned optical design~\cite{sitzmann2018e2e, chang2019diffractive}; $\mathcal{C} = \{\bc_0\}$ recovers adaptive sensor management~\cite{kaelbling1998pomdp, kreucher2005kastella}.

\paragraph*{Two-level structure (joint-DP).}
When $\Pi$ is the full space of adaptive policies, the inner $\argmax_\pi$ is solved by the Bellman recursion over the belief state $\bb_k$.  Bellman's principle of optimality states that every continuation of an optimal policy is itself optimal: at the final step $K$ the optimal action is the one with the best immediate expected reward, and at each earlier step the optimal action is the one whose immediate reward plus the already-known optimal continuation is largest. This backward construction is formalized by the Bellman recursion, which re-writes the inner policy optimization into a belief-dependent, recursive value-to-go functional~\cite{bellman1957dynamic,kaelbling1998pomdp}:
\begin{align}
V_K(\bb;\, \bc) &\;=\; \E_{\bx \mid \bb}\!\left[\,\Phi_0(\bx, \bb;\, \bc)\,\right], \label{eq:bellman_terminal}\\
V_k(\bb;\, \bc) &\;=\; \max_{\bs_k \in \mathcal{S}}\; \E_{\by_k \mid \bb, \bs_k, \bc}\!\left[\,V_{k+1}\!\bigl(T(\bb, \bs_k, \by_k;\, \bc);\, \bc\bigr)\,\right], \label{eq:bellman}
\end{align}
for $k = K{-}1, \ldots, 0$, where $\Phi_0$ is the reward rewritten as a belief-encoded function (see Appendix~\ref{app:phi0}), and 
$T$ is the Bayesian belief-update operator defined in Eq.~\eqref{eq:bayes_update}.  At each step, the recursion selects the action that maximizes the expected value of the updated belief, accounting for the fact that all future actions will also be chosen optimally.  The joint hardware-policy optimum is
\begin{equation}
\bc^\star \;=\; \argmax_{\bc \in \mathcal{C}}\; V^\star(\bc), \qquad V^\star(\bc) \;\coloneqq\; V_0(\bb_0;\, \bc).
\label{eq:jointDP}
\end{equation}

\subsection{The scalar-reward family $\Phi$}
\label{sec:phi_family}

Adaptive policy optimization requires a point-wise reward $\Phi$ that evaluates the inferential outcome against the ground truth $\bx$. Many such rewards may be defined; we consider three specific examples motivated by classical information theory and statistics.

\textit{Mutual-information-aligned reward $\Phi_{\text{MI}}$.}---Following Lindley~\cite{lindley1956measure} through Chaloner--Verdinelli~\cite{chaloner1995bayesian}, this reward targets maximal uncertainty reduction regardless of loss function:
\begin{equation}
  \Phi_{\text{MI}}(\bx, \bs_{1:K}, \bc) \;=\;\ln|\mathrm{supp}(\bb_K)| - H(\bb_K),
\end{equation}
where $H(\bb_K) = -\int \bb_K(\bx) \ln \bb_K(\bx)\, d\bx$ is the differential entropy of the terminal posterior.  Averaging over the prior gives the mutual information $I(\bx; \by_{1:K} \mid \bs, \bc)$.  $\Phi_{\text{MI}}$ is natural for discrete-state problems (where squared error is undefined) and for large-$K$ problems where the posterior concentrates and MI becomes monotone in mean-squared error (MSE) via the Bernstein--von~Mises theorem.

\textit{Bayesian-MSE-aligned reward $\Phi_{\text{Var}}$.}---This reward directly targets squared-error loss:
\begin{equation}
    \Phi_\text{Var}(\bx,\bs_{1:K},\bc)= 
    -\lVert \bx - \hat{\bx}(\bb_K) \rVert^2,
\end{equation}
where $\hat{\bx}(\bb_K)$ is the posterior mean of the terminal belief.
By the tower law of variance,
\begin{equation}
   \E_{\bx,\by_{1:K}}\!\bigl[\, \Phi_\text{Var}(\bx, \bs_{1:K}^{\pi}, \bc) \,\bigr] \;=\; -\E_{\by_{1:K}}\!\bigl[\mathrm{Var}(\bx \mid \bb_K)\bigr],
\end{equation}
for which $V(\bc, \pi)$ equals $-1\times$ the deployed Bayesian MSE exactly, without any asymptotic approximation.  $\Phi_{\text{Var}}$ is the right choice whenever the deployment metric is MSE, and it strictly dominates $\Phi_{\text{MI}}$ as an MSE anchor for continuous-state, finite-$K$ problems with bounded-support priors.

\textit{Fisher-information-aligned reward $\Phi_{\log J}$.}---This reward evaluates the local sensitivity information content at $\bx$ and is suitable for high-dimensional continuous $\bx$:
\begin{equation}
  \Phi_{\log J}(\bx, \bs, \bc) \;=\; \tfrac{1}{2} \log\det J_N(\bx, \bs_{1:K}, \bc),
\end{equation}
where $J_N$ is the accumulated Fisher information matrix~\cite{cramer1946mathematical, rao1945information}. In the Gaussian-posterior regime $\Phi_{\log J}$ agrees with $\Phi_{\text{Var}}$ up to log-transforms, retaining an asymptotic Cramer-Rao bound (CRB) interpretation.

\subsection{Value decomposition and why co-design is necessary}
\label{sec:decomposition}

For any $\Phi$, the outer objective decomposes into three ordered value functionals and a gap:
\begin{align}
V_{\text{oracle}}(\bx, \bc) &= \max_{\bs_{1:K}} \Phi(\bx, \bs_{1:K}, \bc), \label{eq:Voracle}\\
V_{\text{fixed}}(\bc)       &= \max_{\bs_{1:K}} \E_{\bx}\!\bigl[\Phi(\bx, \bs_{1:K}, \bc)\bigr], \label{eq:Vfixed}\\
V_{\text{adaptive}}(\bc)    &= \max_{\pi \in \Pi} \E_{\bx} \E_{\by\mid\bx, \pi, \bc}\!\bigl[\Phi\bigr]. \label{eq:Vadaptive}
\end{align}
The three values are ordered at every $\bc$:
\begin{equation}
V_{\text{fixed}}(\bc) \;\le\; V_{\text{adaptive}}(\bc) \;\le\; \E_{\bx}[V_{\text{oracle}}(\bx, \bc)],
\label{eq:chain}
\end{equation}
because fixed schedules are a subset of adaptive policies (left) and no policy can beat a clairvoyant that knows $\bx$ (right).  Here $V_{\text{fixed}}$ is the expected reward attainable by the best non-adaptive schedule that commits all $K$ actions in advance; $V_{\text{oracle}}$ is the reward a clairvoyant attains by choosing the schedule tailored to each $\bx$; and $V_{\text{adaptive}}$ is the Bellman-optimal adaptive policy, the quantity the user ultimately deploys.

The expected value of perfect information (EVPI), $\E[\mathrm{\Delta_I}](\bc) = \E_{\bx}[V_{\text{oracle}}] - V_{\text{fixed}}$, quantifies the gap between the clairvoyant ceiling and the best blind schedule.  The adaptive gain $V_{\text{adaptive}} - V_{\text{fixed}}$ is bounded above by the EVPI, and this bound motivates a natural question: can maximizing the EVPI over $\bc$ serve as a proxy for the joint-DP optimum?  It cannot.  The EVPI upper bound is slack, and its slackness varies non-monotonically with $\bc$: extreme geometries produce high EVPI because the oracle's advantage is enormous, but the adaptive policy captures only a small fraction of the headroom.  Therefore, joint co-optimization of $(\bc, \pi)$ is necessary:
\begin{equation}
\argmax_{\bc} \,\E[\mathrm{\Delta_I}](\bc) \;\ne\; \argmax_{\bc} V_{\text{adaptive}}(\bc).
\label{eq:argmax_mismatch}
\end{equation}
A geometry that maximizes $V_{\text{fixed}}$ is generically not the geometry that maximizes $V_{\text{adaptive}}$: a geometry good for blind schedules leaves little room for outcome-dependent re-routing, while a geometry with very large oracle value may put most of that value out of reach of any non-clairvoyant policy.  The geometry that deploys well is an \emph{interior} geometry---neither over-tuned to blind schedules nor reliant on clairvoyance.  One cannot pick $\bc$ first and then pick $\pi$, because the criterion for a good $\bc$ depends on which $\pi$ will follow.  On the radar benchmark of Sec.~\ref{sec:radar}, the EVPI-maximizing geometry loses $2.8\times$ in attainable $V_{\text{adaptive}}$.  The EVPI retains value as a diagnostic and pre-screening tool, but not as a design target.

\section{Differentiable dynamic programming for the outer gradient}
\label{sec:envelope}

The outer optimization over $\bc$ requires $dV^\star/d\bc$.  Computing this naively would require differentiating through the Bellman $\argmax$, which is discontinuous.  The dominant approach in differentiable dynamic programming~\cite{mensch2018differentiable, amos2018mpc, domke2012generic, blondel2022efficient} replaces $\max$ with a smooth relaxation (softmax, entropy regularization) so that reverse-mode AD applies everywhere, at the cost of a regularization bias controlled by a smoothing temperature.  We take the alternative path: keep the $\max$ operator sharp and use the \emph{envelope theorem} (Danskin's theorem~\cite{danskin1966theory, milgrom2002envelope}) to certify that at the Bellman-optimal argmax the indirect contribution of argmax sensitivity to $\partial V^\star/\partial \bc$ vanishes exactly.  The resulting gradient is unbiased with no temperature schedule.  To our knowledge, this sharp-max envelope approach has not previously been applied to belief-state POMDPs, where $\bc$ enters both the observation likelihood and the Bayesian belief update $T$.

\subsection{The envelope identity}

Let $\pi^\star(\bc) = \argmax_{\pi} V(\bc, \pi)$.  Differentiating $V^\star(\bc) = V(\bc, \pi^\star(\bc))$,
\begin{equation}
\frac{dV^\star}{d\bc} \;=\; \frac{\partial V}{\partial \bc}\bigg|_{\pi = \pi^\star} \;+\; \underbrace{\frac{\partial V}{\partial \pi}\bigg|_{\pi = \pi^\star}}_{= 0 \text{ at optimum}} \cdot \frac{d\pi^\star}{d\bc}.
\end{equation}
The second term vanishes by first-order optimality~\cite{danskin1966theory, milgrom2002envelope, bonnans2000perturbation}, giving the \emph{envelope identity}:
\begin{equation}
\boxed{\;\frac{dV^\star}{d\bc} \;=\; \frac{\partial V(\bc, \pi^\star(\bc))}{\partial \bc}\bigg|_{\pi^\star \text{ held fixed}}.\;}
\label{eq:envelope_gradient}
\end{equation}
The policy Jacobian $d\pi^\star/d\bc$ does not appear.

\subsection{Unfolding through the Bellman recursion}

The envelope identity reduces the outer gradient to a partial derivative at frozen policy.  Define $Q_k(\bb, \bs_k;\, \bc) \coloneqq \E_{\by_k}[V_{k+1}(T(\bb, \bs_k, \by_k;\, \bc); \bc)]$, so $V_k = \max_{\bs_k} Q_k$.  The per-step envelope identity gives $\partial V_k / \partial \bc = \partial Q_k / \partial \bc |_{\bs_k^\star}$, which unfolds as
\begin{align}
\frac{\partial V_k}{\partial \bc} \;=\; \E_{\by_k}\!\Biggl[
&\;\frac{\partial \log p(\by_k \mid \bb, \bs_k^\star, \bc)}{\partial \bc}\, V_{k+1} \notag \\
& + \frac{\partial V_{k+1}}{\partial \bb} \cdot \frac{\partial T}{\partial \bc} + \frac{\partial V_{k+1}}{\partial \bc}\;\Biggr],
\label{eq:dV_unfolded}
\end{align}
with three terms: (i)~how the measurement probability itself changes with $\bc$ (log-likelihood score term), (ii)~how the Bayesian belief update shifts when $\bc$ changes the likelihood, and (iii)~the recursive contribution from all future steps.

Operationally, we can evaluate the gradients by judiciously leveraging automatic differentiation (AD). (1)~Run backward induction from $k=K$ to $k=0$, populating $V_k$ and recording $\bs_k^\star$ at every reachable belief.
(2)~Re-evaluate $V_0(p_0; \bc)$ substituting the recorded $\bs_k^\star$ in place of $\max_{\bs_k}$; the resulting expression is argmax-free.
(3)~Apply reverse-mode AD; the expression is composed of closed-form $\bc$-dependent primitives, and AD returns $\partial V_0/\partial \bc$ exactly.
The natural outer loop alternates Bellman solves with gradient steps:
(1)~initialize $\bc^{(0)}$;
(2)~solve $\pi^\star(\bc^{(t-1)})$ by inner DP;
(3)~update $\bc^{(t)} = \bc^{(t-1)} + \eta \cdot \partial V / \partial \bc$;
(4)~box-project $\bc^{(t)}$ onto the feasible set; repeat until convergence.

\section{A hierarchy of relaxations}
\label{sec:hierarchy}

Exact Bellman DP is tractable only for small POMDPs.  We propose a ladder of principled relaxations where each rung replaces the Bellman optimum with a \emph{plausibe policy}: each rung prescribes $\pi$ in closed form as a principled heuristic, and the outer loop moves $\bc$. 

\paragraph*{Rung~1: Exact Bellman DP.}
The policy $\pi^\star_k = \argmax_{\bs_k} Q^\star_k(\bb_k, \bs_k)$ is the true Bellman optimum, with the exact belief $\bb_k$ tabulated on a grid (continuous state) or stored as a probability vector (discrete state).  The feasibility ceiling is set by the reachable-belief memoization: approximately $|\mathcal{S}|^K \lesssim 10^6$ nodes on a single processor thread.  Our case studies~A and~B operate at this rung. While this paper establishes the conceptual foundation of joint-DP by demonstrating co-design with small POMDPs, future works in this rung will systematically pursue
the Bellman optimum for high-dimensional co-design by devising ``outer-aware'' and scalable approximate DP algorithms.

\paragraph*{Rung~2: Certainty-equivalent control.}
At step $k$, form the posterior mean $\hat\bx_k = \E[\bx \mid \bb_k]$ and play the schedule the clairvoyant oracle would choose against $\hat\bx_k$.  This is near-optimal when the posterior concentrates rapidly and the oracle map is Lipschitz in $\bx$, but it can under-explore when $\hat\bx_k$ is wrong.

\paragraph*{Rung~3: Myopic one-step.}
Collapse the look-ahead to one step: $\bs_k = \argmax_{\bs} \E_{\by_k \mid \bb_k, \bs}[\Phi(\bx, \bs, \bc)]$ while recursive Bayesian updates are computed by differentiable particle filters.  This is justified by submodularity bounds~\cite{drovandi2014sequential} for information-theoretic rewards but misses multi-step payoffs.

\paragraph*{Rung~4: per-step Fisher-maximization with EKF.}
Replace the full posterior by Gaussian moments $\bb_k \approx \mathcal{N}(\bar\bx_k, \Sigma_k)$ propagated by an extended Kalman filter (EKF), and select $\bs_k = \argmax_{\bs} \tr J(\bar\bx_k, \bs, \bc)$.  This is the computationally cheapest rung compatible with continuous $\bc$ and arbitrary state dimensions.  The Gaussian assumption and EKF may fail under strongly multi-modal posteriors but may be replaced by parametric multi-modal distributions and filters. Our case study~C operates at this rung.

Each rung trades Bellman optimality for scalability while maintaining implicitly defined inner policies: rung~1 is exact but memory-limited; rung~2 replaces the Bellman lookahead with the oracle at a point estimate; rung~3 maximizes single-step reward without planning ahead; and rung~4 replaces the full posterior with Gaussian moments.  When one rung's feasibility ceiling is exceeded, the practitioner descends to the next.

\paragraph*{Dichotomy at the exact-DP branch point.}  Our co-design framework evidently exhibits strong affinity towards reinforcement learning (RL), which may be directly leveraged to replace the inner Bellman recursion. The dominant approach in modern RL is to parametrize the policy $\pi_\theta$ (or the action-value $Q_\theta$) with a neural network and optimize it by policy gradient, $Q$-learning, or actor--critic~\cite{hausknecht2015deep, zhong2018deep}.  \emph{We do not pursue this branch in the present paper, though we do not rule it out for future work.}  The reason is a well-documented pathology: an over-expressive neural-network (NN) policy can absorb the training signal and compensate for a poor $\bc$ by investing in policy complexity, leaving $\bc$ near its initialization.  The outer loop converges to a \emph{hardware-light} fixed point that looks good in simulation but relies on a near-trivial $\bc$ hard to justify as a physical design.  Known countermeasures partially mitigate the risk but do not yet admit the clean sharp-max differentiable-DP gradient our framework exploits: capacity control on $\pi_\theta$ (bottleneck width, weight decay, dropout); two-timescale updates with $\bc$ stepped on a faster clock than $\theta$; warm-starts from a prescribed inner rule (such as rungs 3--4); and auxiliary regularizers penalizing hardware-null geometries.  We flag the combination of these remedies as a plausible path to extending joint-$(\bc, \pi)$ to expressive NN policies in future work, and correspondingly exclude NN-parametrized policies from our relaxation ladder on a \emph{provisional} rather than principled basis.

\section{Case Study A: Radar beam search}
\label{sec:radar}

\begin{figure}[t]
\centering
\includegraphics[width=\columnwidth]{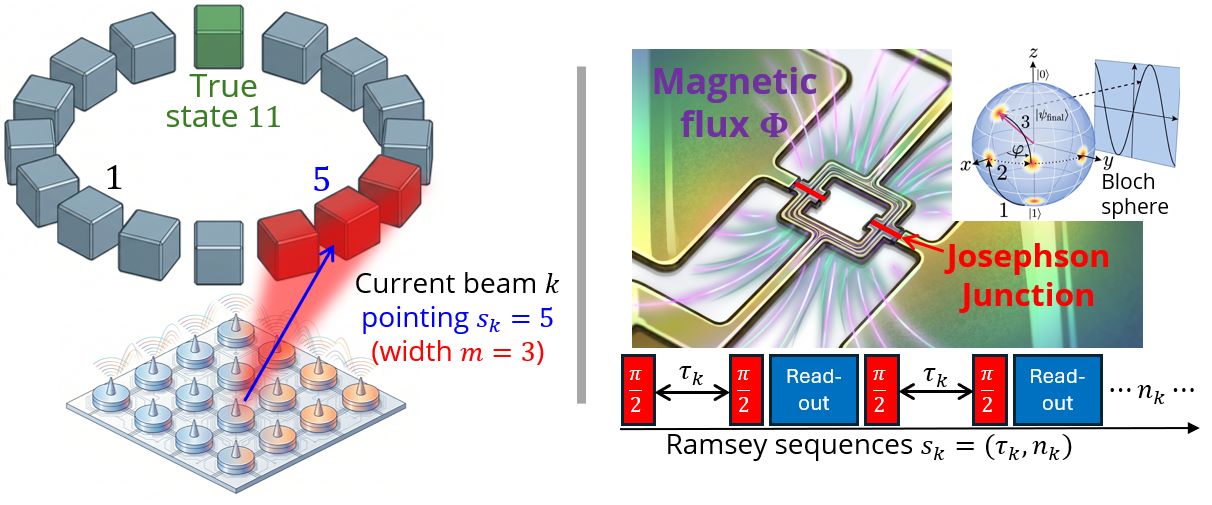}
\caption{Scene illustrations.  Left: Case~A radar POMDP---a phased-array antenna interrogates a ring of 16 target cells.  Right: Case~B superconducting-qubit flux sensor---x-mon transmon with SQUID loop, external flux $\Phi_\mathrm{ext}$, and adaptive Ramsey sequences.}
\label{fig:casesAB}
\end{figure}

This case study applies the joint-DP framework to a pedagogical \emph{toy} problem: a discrete radar beam search where exhaustive enumeration of all measurement-observation sequences is tractable. Because the design variable (beamwidth) is a discrete integer that can be swept, this case does not require computing the outer gradient of Eq.\,\eqref{eq:dV_unfolded} and instead serves to illustrate the value decomposition.

\subsection{Problem setup}

The inference task is to localize a target on a 16-cell ring using a sensor with a tunable beamwidth over $K = 4$ measurement steps.  The discrete state, observation, and design spaces together allow exact computation of all three value functionals across the design sweep.

\paragraph*{State.} The state $\bx \in \{1, \ldots, \chi\}$ ($\chi=16$) indexes the target's cell, with uniform prior $p_0(\bx) = 1/\chi$.

\paragraph*{Action.}  At each step $k$, the sensor with fixed odd-integer beamwidth $m \in \{1, 3, \ldots, \chi-1\}$ is directed at cell $\bs_k \in \{1, \ldots, \chi\}$, covering cells within ring distance $m/2$ of $\bs_k$. Inside the beam, the detection probability is $p_\mathrm{D} = 0.9$; outside, the false-alarm probability is $p_\mathrm{FA} = 0.05$.

\paragraph*{Observation.} The observation $\by_k \in \{0, 1\}$ indicates detection ($\by_k = 1$) or no-detection ($\by_k = 0$), with Bernoulli likelihood
\begin{equation}
\begin{split}
p(\by_k = 1 \mid \bx, \bs_k; m) &= p_\mathrm{FA} \\
&\quad + (p_\mathrm{D} - p_\mathrm{FA})\, G_m(\theta_{\bs_k} - \varphi_{\bx}),
\end{split}
\end{equation}
where $\varphi_{\bx} = 2\pi(\bx - 1)/\chi$ is the bearing of cell $\bx$, $\theta_{\bs_k}$ is the pointing angle, and $G_m$ is the top-hat aperture of angular full-width $2\pi m/\chi$ radians.

\paragraph*{Reward.}  The radar search is an identification task rather than continuous estimation, so a suitable reward functional is the terminal mutual information, $\Phi = \Phi_\text{MI}$.

\begin{figure}[t]
\centering
\hspace*{-0.6cm}
\includegraphics[width=0.52\textwidth]{figures/CaseA_radar_v3.png}
\caption{Case Study~A: Radar Beam Search. (a)~Mean terminal ($k=4$) MI (solid) and TG (dashed, coinciding for adaptive and fixed) under the adaptive policy (green), oracle schedule (blue), and fixed schedule (red) as a function of beamwidth $m$. (b)~Belief evolution over measurement steps $k=1,\ldots,4$ at saturated beamwidth $m=9$, under the adaptive policy (left columns) and fixed schedule (right columns). Black arrow: beam center. Shaded sector: beamwidth coverage; green on detection, red otherwise. Cell~8 (black) is the target. Grey radial bars: posterior after each measurement. (c),~(d)~Scatter plot of the joint terminal (MI, TG) distributions in the saturated regime (c, $m=9$) and unsaturated regime (d, $m=3$), with marker area proportional to likelihood (omitted below $0.1\%$). Quadrants label inference quality: confidently (MI$>1$) or broadly (MI$<1$) correct (TG$>0$) or incorrect (TG$<0$).}
\label{fig:radar_results}
\end{figure}

\subsection{Adaptive structure at the joint-DP optimum}

\begin{table}[h]
\caption{Value functionals for selected beam widths (nats).  Bold: column maximum.}
\label{tab:radar}
\begin{ruledtabular}
\begin{tabular}{rrrrrr}
$m$ & $V_\text{oracle}$ & $V_\text{adaptive}$ & $V_\text{fixed}$ & $\E[\Delta_I]$ & Sat.\ \% \\
\hline
 1 & 2.580 & 0.619 & 0.499 & \textbf{2.081} & 5.8 \\
 5 & 2.081 & 1.631 & 1.343 & 0.737 & 38.9 \\
 7 & 1.924 & 1.700 & \textbf{1.411} & 0.513 & 56.4 \\
\textbf{9} & 1.879 & \textbf{1.704} & 1.401 & 0.478 & \textbf{63.5} \\
13 & 2.312 & 1.389 & 1.041 & 1.271 & 27.3 \\
\end{tabular}
\end{ruledtabular}
\end{table}

The value functionals $V_\text{oracle}$, $V_\text{adaptive}$, and $V_\text{fixed}$ across beamwidths are shown in Fig.~\ref{fig:radar_results}(a) and Table~\ref{tab:radar}.  The joint-DP optimum is at $m^\star = 9$ ($V_\text{adaptive}^\star = 1.704$ nats).  The ignorance gap $\E[\mathrm{\Delta_I}] = V_\text{oracle} - V_\text{fixed}$ is non-monotone in $m$: it peaks at the two extremes ($m = 1$ and $m = 15$) and reaches its minimum at $m^\star = 9$, coincident with the $V_\text{adaptive}$ peak.  Selecting geometry by EVPI alone would push toward the extremes and lose $2.75\times$ in attainable $V_\text{adaptive}$ relative to the joint solve.

At saturated beamwidths, the Bellman-optimal policy emulates a binary search, bisecting the current posterior at each step.  Action-observation sequences, using $m=m^\star$ with $\bx=8$, are shown in Fig.~\ref{fig:radar_results}(b) for the optimal fixed schedule and optimal adaptive policy. Under the adaptive policy (left), successive non-detections at cells 1 and 14 narrow the support before detections at cells 12 and 4 localize the target. By contrast, the fixed schedule (right) must commit its full pointing sequence in advance. The initial detection coarsely localizes the posterior, but subsequent pre-committed pointings cannot reliably bisect what remains, and the terminal belief lands split between the target (cell 8) and an adjacent cell (cell 7).
\vspace*{.2\baselineskip}
\subsection{Decisive inference in the saturated regime}

The action-observation sequences in Fig.~\ref{fig:radar_results}(b) illustrate the adaptive-fixed contrast for a single target placement. To characterize the adaptive policy's advantage statistically, we complement MI with the truth gain $\text{TG}(b, \bx) = \ln\bigl(\chi\, b(\bx)\bigr)$, which distinguishes confidently correct posteriors from confidently incorrect ones: TG is zero at the uniform prior, $\ln \chi$ when belief is a delta on the true cell, and negative when belief puts less mass on truth than uniform.  The mean terminal TG coincides with the mean terminal MI for the adaptive policy and fixed schedule [Fig.~\ref{fig:radar_results}(a), dashed lines], so any difference between the two must reside in the joint distribution rather than the mean.

The joint (MI, TG) distributions are shown in Fig.~\ref{fig:radar_results}(c, d) for saturated ($m^\star = 9$) and unsaturated ($m = 3$) beamwidths, with outcomes partitioned into four quadrants (confidently/broadly $\times$ correct/incorrect).  In Fig.~\ref{fig:radar_results}(c), the adaptive distribution concentrates in the high-MI quadrants, with $85\%$ confidently correct realizations (MI $> 1$, TG $> 0$) versus $74\%$ for the fixed schedule, an 11-percentage-point gain in actionable, correctly-localized outcomes.  The contrast is sharpest in the broadly-correct quadrant (MI $< 1$, TG $> 0$), where the fixed schedule lands $12\%$ of the time on diffuse-but-correct posteriors like those illustrated in Fig.~\ref{fig:radar_results}(b) (right), while the adaptive policy reduces this fraction to $3\%$.

The observation-conditioned re-planning of the adaptive policy leads to decisive inference by bisecting the current posterior at each step.  However, the bisection proceeds regardless of observation-noise events (missed detections or false alarms), driving broad-MI fates near zero regardless of TG sign ($3\%$ correct, $0.1\%$ incorrect).  The fixed schedule, lacking re-planning, retains lower-MI modes.  At unsaturated beamwidth $m = 3$ [Fig.~\ref{fig:radar_results}(d)], both adaptive and fixed distributions disperse across a wide MI range; confidently correct adaptive sequences are contingent on favorable early detection (no detection) for narrow (wide) beamwidths.  Notably, the probability of incorrect inference is nearly conserved between the saturated and unsaturated configurations ($11.9\%$ at $m^\star$ vs $12.8\%$ at $m = 3$).  The confidently incorrect tail at $m^\star$ is a residual of detector noise that shrinks with $p_\text{D} \to 1$ and $p_\text{FA} \to 0$.

\section{Case Study B: Superconducting-qubit flux sensor}
\label{sec:scqubit}

This case study takes the same exact-DP machinery to a continuous-state problem whose deployment metric is Bayesian MSE, and compares joint-DP co-design against the joint Bayesian Cram\'er--Rao fixed-schedule baseline that dominates the classical Fisher-information literature for scalar estimation.

\begin{figure}[t]
\centering
\includegraphics[width=\columnwidth]{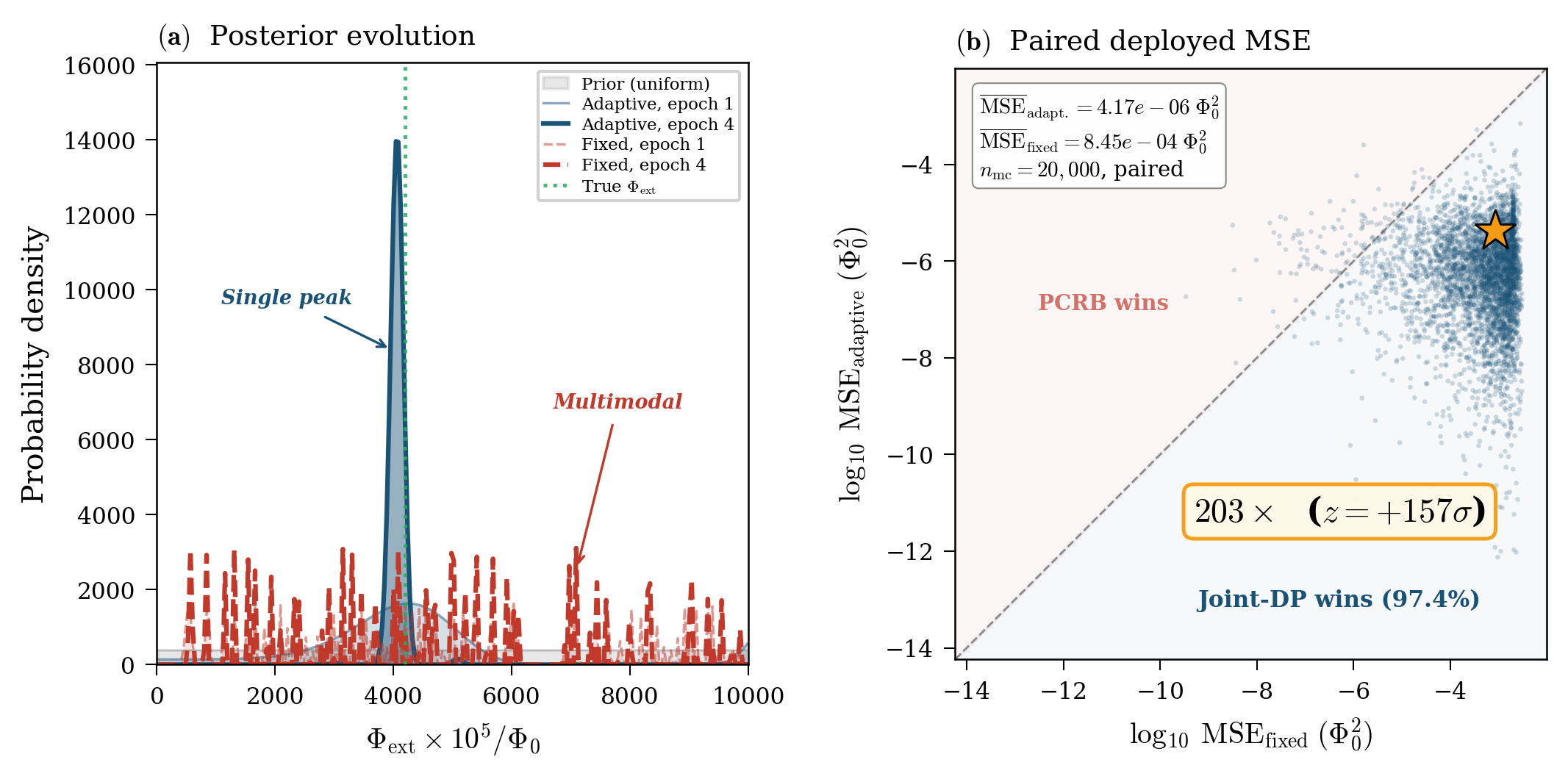}
\caption{Case~B results.  (a)~Posterior evolution for a single trajectory at $\Phi_\mathrm{ext} = 0.042\,\Phi_0$: the adaptive policy (blue) collapses to a single peak after epoch~1, while the fixed schedule (red dashed) remains multimodal.  (b)~Paired deployed MSE ($n_\mathrm{mc} = 20{,}000$, same $\Phi_\mathrm{true}$ per trajectory): each dot is one trajectory, the diagonal is break-even.  Joint-DP wins on $>$97\% of trajectories with a $202\times$ mean ratio at $z = +157\sigma$.}
\label{fig:scqubit_results}
\end{figure}

\subsection{Problem setup}

A frequency-tunable transmon sensor~\cite{danilin2024scqubit} measures external magnetic flux $\bx = \Phi_\mathrm{ext}$ threading a SQUID loop.  The five free hardware parameters are $\bc = (f_q^{\max}, E_C/h, \kappa, \Delta_{qr}, \omega_d)$, with noise amplitudes and temperature held fixed at the operating values reported in Ref.~\cite{danilin2024scqubit}.

\paragraph*{State.}  The unknown is the dimensionless flux $\phi \equiv \Phi_\mathrm{ext}/\Phi_0$, with prior $\phi \sim \mathrm{Uniform}[0, \phi_{\max}]$ at $\phi_{\max} = 0.1\,\Phi_0$.

\paragraph*{Action.}  At each of $K = 4$ epochs the policy selects a Ramsey delay $\tau_k$ from a grid of $J = 10$ log-spaced values in $[10, 320]$~ns and a repetition count $n_k \in \{1, 10\}$, giving the action $\bs_k = (\tau_k, n_k)$.

\paragraph*{Observation.}  Ramsey interferometry returns a Bernoulli outcome (Binomial when $n_k > 1$) with likelihood
\begin{equation}
p(\by{=}1 \mid \bx, \tau; \bc) = \tfrac{1}{2} + \tfrac{1}{2}\,\mathrm{th}(\bx, \bc)\,e^{-A\tau - B^2 \tau^2}\,\cos[\Delta\omega\, \tau],
\label{eq:ramsey}
\end{equation}
with all noise rates and spectral coefficients closed-form in the parametric geometry $\bc$.

\paragraph*{Reward.}  The deployment metric is Bayesian MSE, so $\Phi = \Phi_{\text{Var}}$, which makes $V(\bc, \pi)$ equal to $-1\times$ the deployed MSE exactly (Sec.~\ref{sec:phi_family}).

The Ramsey likelihood is invariant under permutation of measurements at the same $(\tau, n)$ setting, so the belief $\bb_k$ is a sufficient statistic of the cumulative counts $(N_j, M_j)_{j=1}^J$ (total measurements and detections at each delay).  The Bellman DP therefore admits exact memoization on a count-tuple hash table; at the parameters above the table contains $\sim 10^7$ nodes and the DP completes in minutes on a single CPU thread.

\subsection{Joint-DP co-design and comparison with PCRB baseline}

We optimize the five hardware parameters over a physically realistic regime: ($f_q^{\max} \in [3, 12]$~GHz, $E_C/h \in [0.15, 0.4]$~GHz, $\kappa \in [0.1, 5]$~MHz, $\Delta_{qr} \in [0.8, 5]$~GHz, $\omega_d/(2\pi) \in [1, 12]$~GHz), using multi-start gradient-based MMA optimizer (multi-start to exhaustively search the 5D design space since gradient methods are prone to local minima in a low-dimensional design space).  The optimized geometry is
\begin{equation}
\bc^\star_1 \approx (3.00\;\text{GHz},\; 0.15\;\text{GHz},\; 0.48\;\text{MHz},\; 4.02\;\text{GHz}),
\end{equation}
with $\omega_d/(2\pi) = 1.00$~GHz.  All five parameters sit at or near the lower bounds of the feasible box.

The posterior Cramer-Rao bound (PCRB) gives an alternative single-level optimization: $(\bc^\star_2, \bs^\star_2) = \argmax_{\bc,\, \bs_{1:K}} \log J_P(\bs_{1:K}, \bc)$, yielding $\bc^\star_2 \approx (12.0\;\text{GHz},\; 0.4\;\text{GHz},\; 0.10\;\text{MHz},\; 5.0\;\text{GHz})$ with $\omega_d/(2\pi) = 1.00$~GHz and fixed schedule $(\tau_{10}, n_2)^4 = (320\,\text{ns}, 10)^4$.  The two optimizers converge to \emph{physically opposite corners} of the feasible box: joint-DP at the lower bounds of $f_q^{\max}$ and $E_C/h$, PCRB at their upper bounds.  Both independently select $\omega_d$ at the lower bound, far below the conventional single-shot-resonant heuristic---off-resonant drive turns out to be optimal for both objectives. Deploying each design with $n_{\mathrm{mc}} = 20{,}000$ paired Monte-Carlo trajectories (common $\Phi_{\text{true}}$ draws) gives the results in Fig.~\ref{fig:scqubit_results}(b).  Joint-DP beats the PCRB baseline by $202\times$ on deployed Bayesian MSE.

The physical origin of the gap is visible in the posterior evolution of Fig.~\ref{fig:scqubit_results}(a).  The Fisher information at long Ramsey delay may become high because one long measurement gives sharp local phase resolution, but at prior width $\phi_{\max} = 0.1\,\Phi_0$ the prior spans many fringes and the single-level Fisher criterion has no machinery for disambiguating which fringe the true $\Phi_\mathrm{ext}$ lies on.  The Bellman-optimal policy resolves this: a short-$\tau$ first epoch collapses the multimodal posterior to a unimodal one, and subsequent epochs exploit long-$\tau$ resolution within the selected fringe.  The PCRB-optimal schedule, which commits to $\tau = 320$~ns from step~1, cannot undo the resulting aliasing.  Moreover, the $202\times$ MSE gap is not a consequence of a better algorithm applied to the same hardware; it is a consequence of different hardware designed with the algorithm's capabilities in mind---the joint-DP optimizer selects a low-$f_q^{\max}$, high-$\Delta_{qr}$ geometry that produces fast Ramsey fringes a fixed schedule cannot disambiguate but the adaptive policy can.

\subsection{Aliasing-resistant adaptive structure}

The adaptive advantage described above is intimately connected to Ramsey aliasing.  The prior width $\phi_{\max}$ interacts with the tractable horizon $K = 4$ through the periodicity of the Ramsey likelihood in $\Delta\omega(\bx)\,\tau$: a prior that spans many fringes yields a multimodal posterior that can be disambiguated only by measurements at sufficiently varied $\tau$.  At $\phi_{\max} \to \Phi_0/2$ both methods converge to comparable aliasing-floor MSE because four epochs are too few for either a fixed or an adaptive schedule to resolve the many fringes.  At $\phi_{\max}$ well below the aliasing threshold both operate far below the floor and the margin grows rapidly.  Choosing $\phi_{\max} = 0.1\,\Phi_0$ sits just inside the regime where the adaptive finesse becomes observable at $K = 4$ while keeping the PCRB baseline valid.  The adaptive policy's anti-aliasing strategy is visible in Fig.~\ref{fig:scqubit_results}(a): the first epoch at short $\tau$ narrows the posterior support from many fringes to one, and the remaining epochs resolve within that fringe at long $\tau$---a coarse-to-fine strategy that the fixed schedule, having committed all four epochs to long $\tau$, cannot replicate.

\section{Case Study C: Photonic metasensor}
\label{sec:photonic}
While sharp-max DP handles the inner Bellman exactly on low-dimensional belief states, scaling the outer design loop to topology optimization (with $10^3-10^9$ design pixels) with full-wave Maxwell solves demands an extensive relaxation of the inner problem. As a preliminary step toward scalable co-design, we relaxed the Bellman policy into an information-theoretic greedy policy (per-step Bayesian-Fisher-maximization) operating on the running extended-Kalman-filter (EKF) posteriors (rung 4). We then coupled this policy to full-wave Maxwell simulations to topology-optimize a proof-of-concept adaptive photonic sensor. The end-to-end gradient becomes the composition of an implicit-differentiation pullback of the inner policy optimum with frequency-domain adjoint analysis. 


\begin{figure}[t]
\centering
\includegraphics[width=\columnwidth]{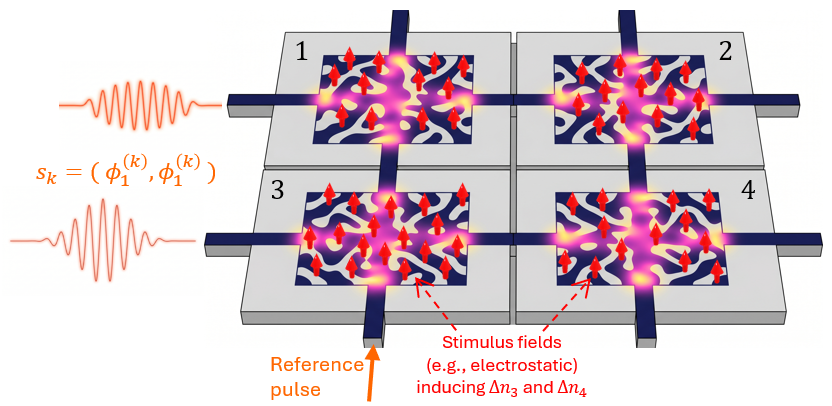}
\caption{Case~C: A $2 \times 2$ lattice of cross-waveguide scatterers with inverse-designed binary permittivity.  Two broadband pulses probe the device; 8 output port powers form $\by_k \in \R^8$; an EKF reconstructs per-cell refractive-index shifts.}
\label{fig:photonic_scene}
\end{figure}

\subsection{Problem setup}

The device is a $2 \times 2$ lattice of identical 2D cross-waveguide scatterers (Fig.~\ref{fig:photonic_scene}), each with a $300 \times 300$-pixel permittivity design region; all cells share the same inverse-designed geometry, giving $\bc \in \R^{90{,}000}$ as the outer optimization variable.

\paragraph*{State.}  The unknown $\bx = \mathrm{vec}(\Delta n) \in \R^4$ is the vector of per-cell refractive-index perturbations.

\paragraph*{Action.}  At each of $K = 3$ epochs the sensor chooses a pair of input phases $\bs_k = (\phi_1^{(k)}, \phi_2^{(k)})$ applied to three Gaussian-pulse probes.

\paragraph*{Observation.}  The observations $\by_k \in \R^8$ are broadband port powers measured at the lattice's 8 external output ports, with additive Gaussian noise.

\paragraph*{Reward.}  The deployment metric is Bayesian MSE, so $\Phi = \Phi_{\text{Var}}$, with the EKF providing the terminal Gaussian posterior $\bb_K \approx \mathcal{N}(\bar\bx_K, \Sigma_K)$.  The intractable Bellman inner $\argmax$ is replaced by the rung-4 surrogate: at each epoch the prescribed policy maximizes the Bayesian Fisher-information trace at the EKF posterior mean, $\bs^\star_k = \argmax_{\bs} \tr J_k(\bar\bx_k, \bs, \bc)$, solved by L-BFGS optimizer.  The derivative of this inner argmax with respect to $\bc$ is computed via the implicit function theorem.

\subsection{Physics and differentiability}

The forward model is a 2D finite-difference frequency-domain (FDFD) Maxwell solver returning one $4 \times 4$ $S$-matrix per frequency.  To avoid re-running the full FDFD solve at every EKF step, the per-cell $S$-matrix is Taylor-expanded to second order in the local refractive-index perturbation $\Delta n$: $S(\Delta n) \approx S_0 + S'\Delta n + \tfrac{1}{2}S''\Delta n^2$.  The Taylor coefficients $S_0$, $S'$, $S''$ are computed once per outer iterate via a single FDFD solve per frequency, and the four cells are interconnected by port-connection equations to assemble the lattice's 8-port $S$-matrix.  The Jacobian $\partial \by / \partial \bc$ passes through these coefficients via the FDFD adjoint method~\cite{hughes2018adjoint, minkov2020inverse}.  Density filtering with $\tanh$ projection~\cite{wang2011projection, sigmund2007morphology, lazarov2016length, christiansen2018compact} ensures fabrication compatibility ($\beta$ continuation $16 \to 256$).

\subsection{Results and adaptive policy structure}

\begin{figure}[t]
\centering
\includegraphics[width=\columnwidth]{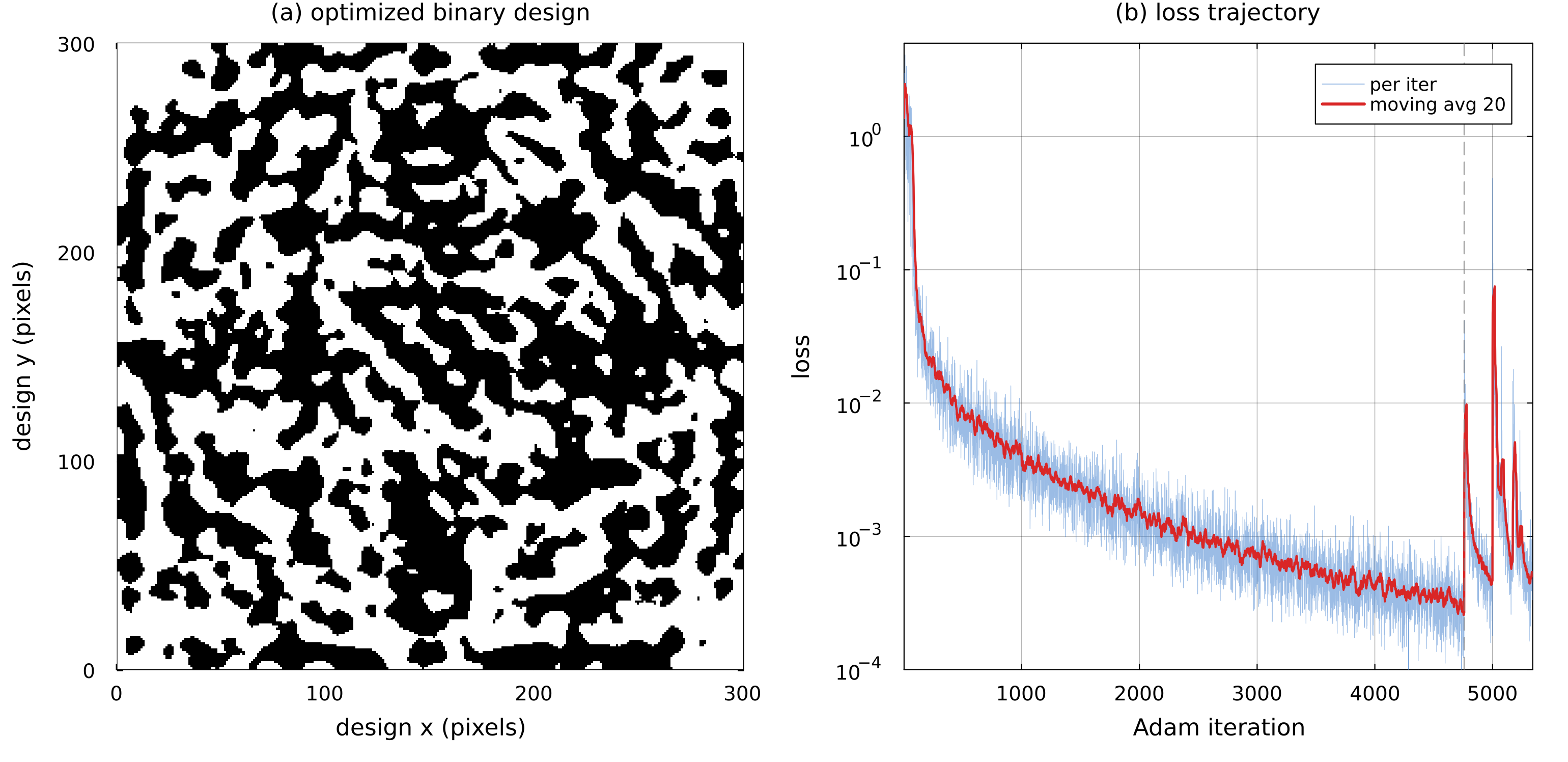}
\caption{Optimized binary permittivity (left) and loss trajectory (right).}
\label{fig:photonic_main}
\end{figure}

The optimized geometry (Fig.~\ref{fig:photonic_main}) reduces posterior-mean relative error of a randomized baseline from 120\% to 0.97\% after ten EKF steps, a $123\times$ improvement; after a single broadband measurement, the EKF reduces relative error from 184\% (prior) to 3.4\%, which is further reduced to 0.97\% ($\sim 4\times$) in subsequent adaptive measurements.  The final binary permittivity is smooth at the density-filter length scale ($R = 5$ pixels $\approx 200$~nm) and compatible with electron-beam lithography.  At the design wavelength $\lambda = 1550$~nm, the shot-noise-limited detection threshold corresponds to $\sim 64$~$\mu$W per port at $1$~$\mu$s integration, within the operating range of standard telecom lasers.

The rung-4 adaptive policy re-selects the input phases $\bs_{k+1}$ after each measurement so that the next probe is aligned with the largest remaining eigenvector of the EKF posterior covariance $\Sigma_k$.  Successive measurements therefore shrink the least-constrained direction of the state estimate first, concentrating the information budget where it is most needed.  This eigenvector-tracking strategy is the continuous-state analog of the binary-search bisection seen in Case~A: both direct successive measurements toward the part of the state space that remains most uncertain.

\section{Discussion and Conclusion}
\label{sec:discussion}

\subsection{The adaptive advantage and what co-design buys}

Across the three case studies the adaptive policy's margin comes from one structural move: redirecting the next measurement based on whichever partition of the state space the current data has already ruled out.  In Case~A, the adaptive policy emulates a binary search conditioned on each detection outcome, bisecting the posterior support at every step.  In Case~B, the first epoch at short $\tau$ collapses the multimodal posterior before later epochs switch to long $\tau$ to sharpen within a single fringe.  In Case~C, each new pair of probe phases is re-aligned with the largest remaining eigenvector of the EKF posterior covariance, so that successive measurements shrink the least-constrained direction of the state.

Information loss at the analog-to-digital boundary upper-bounds any downstream reconstruction.  Joint $(\bc, \pi)$ optimization responds by shaping the physical geometry so that the hardware itself participates in the inference rather than merely transducing for it.  The joint-DP margin is therefore not a numerical artifact of any particular problem: it is the signature of a hardware operating point chosen in anticipation of a conditional policy, and hardware chosen without that anticipation cannot, in general, recover it by attaching an adaptive policy after the fact.  

\subsection{Limitations and open directions}

\textit{Horizon.}---Cases~A and B operate at $K = 4$ where exact DP is tractable; at larger horizons the memoization blows up. Future works will pursue principled approximations of Bellman's recursion that are also ``outer-aware,'' e.g., new co-design algorithms that judiciously amortize the intractable belief state exploration across the stochastic gradient descent campaign of the outer geometry.

\textit{Finite-sample bounds.}---The PCRB is known to be loose relative to the MSE in adaptive sensing problems~\cite{tsang2016quantum, reginatto2019crb}. Future works will investigate computationally more expensive adaptation-oriented bounds such as the Weiss--Weinstein~\cite{weiss1988lower} and Ziv--Zakai~\cite{tsang2017ziv} bounds.

\textit{Dynamical state.}---All three case studies treat $\bx$ as static.  Dynamical $\bx$ (moving targets, drifting parameters) requires recursive predict-correct belief updates, and dynamics-aware Bellman backups. Furthermore, complex dynamical systems naturally invite intervention and control, calling for unified co-optimization of physics, geometry, policy, inference, and control---a natural next step towards co-designed physical intelligence.  

\subsection{Summary and Outlook}
\label{sec:conclusion}
When a sensor's physical hardware is designed once but its measurement policy runs for the device's lifetime---programmable metasurfaces, cognitive radars, frequency-tunable quantum sensors, reconfigurable photonic meshes---joint optimization of geometry and policy is the principled procedure to attain the formally optimal co-design.  Sharp-max differentiable dynamic programming, with the envelope theorem as the exactness certificate, makes the joint optimization tractable via an unbiased outer gradient; a family of compatible scalar rewards makes it applicable across communities; and a principled relaxation hierarchy makes it portable across scales.  The three case studies provide evidence for the prowess of this approach. In each case, the margin is not a consequence of a better algorithm applied to the same hardware, but of different hardware designed with the adaptive policy's capabilities in mind.

\paragraph*{Intelligence migration from digital to physical.}
Information loss at the analog-to-digital boundary upper-bounds any downstream reconstruction; joint $(\bc, \pi)$ optimization responds by shaping the physical geometry to participate in the inference rather than merely transduce for it.  A co-optimized $\bc$ is no longer a lens, an antenna, or a SQUID loop that merely transduces; it is a purpose-built information-preserving structure that filters, concentrates, and encodes the incoming signal in the geometry's own complex degrees of freedom.

\paragraph*{A generalization to RL.} Our geometry-policy co-design framework has a natural affinity with reinforcement learning, but a few distinctions are in order. Standard RL and modern embodied policies map observations directly to actions on a fixed sensing-and-actuation body: a fixed robot, a fixed sensor, a fixed actuator. They neither maintain an explicit belief over the latent state nor select actions to actively reduce its uncertainty, let alone reshape the hardware that determines what is observable in the first place. RL thus designs the digital brain, the policy network, for a body it takes as given. Our approach inverts the question. Before the brain comes the body it inhabits: the photonic substrate that determines what is even sensible, what is even controllable, what is even computable. Extended from sensing through actuation to control, co-design becomes a paradigm that seeks the best body to nurture the best possible brain, dissolving the body/brain distinction as the physical structure becomes an inseparable participant in the inference and control loop. The endgame is not a smart algorithm in a dumb body, but a co-designed substrate-and-cognition in which the two are no longer separable.

\begin{acknowledgments}
This work is supported by the U.S.\ Army Research Office (award numbers W911NF2410390 and W911NF2510113).
\end{acknowledgments}

Data and code underlying the results presented in this paper may be obtained from the authors upon reasonable request.

\appendix
\section{Belief-encoded reward}
\label{app:phi0}
The original reward $\Phi$ takes the action trajectory $\bs_{1:K}$ as an argument. The POMDP Bellman recursion operates on belief-dependent value-to-go functionals. Therefore, we want to replace the action trajectory $\bs_{1:K}$ by the terminal belief $\bb_K$. The substitution is legitimate iff $\Phi$'s dependence on $\bs_{1:K}$ actually factors through $\bb_K$.

\emph{Definition (belief-encodable reward).}  We call $\Phi$ \emph{belief-encodable} if there is a function $\Phi_0(\bx, \bb;\, \bc)$ such that, for every $\bx$, every $\bc$, and every pair of trajectories $(\bs_{1:K}, \by_{1:K})$ and $(\bs'_{1:K}, \by'_{1:K})$ whose terminal beliefs agree ($\bb_K = \bb'_K$),
\begin{equation}
\Phi(\bx, \bs_{1:K};\, \bc) \;=\; \Phi(\bx, \bs'_{1:K};\, \bc) \;=\; \Phi_0(\bx, \bb_K;\, \bc).
\label{eq:belief_encodable}
\end{equation}
In words: any two histories that land on the same terminal posterior yield the same reward.  If Equation~\eqref{eq:belief_encodable} holds, $\Phi_0$ is uniquely defined (up to values on unreachable beliefs), and substituting $\Phi$ with $\Phi_0$ for Bellman recursion is valid.

Two of the rewards considered in this paper pass the test immediately, because each is a simple functional of the posterior's shape:

\begin{itemize}[nosep, leftmargin=*]
\item \textbf{Posterior variance ($\Phi_{\text{Var}}$).}  $\Phi(\bx, \bs_{1:K};\, \bc) = -(\bx - \hat{x}(\bb_K))^2$, where $\hat{x}(\bb_K) = \E[\bx \mid \bb_K]$ is the posterior mean.  The only place $\bs_{1:K}$ enters is through $\bb_K$, and $\bb_K$ is itself the object we are substituting in. Therefore $\Phi_\text{Var}$ is straightforwardly belief-encodable with $\Phi_0(\bx, \bb;\, \bc) = -(\bx - \hat{x}(\bb))^2$.  The belief-level average $\E_{\bx \mid \bb}[\Phi_0]$ equals $-\mathrm{Var}(\bx \mid \bb)$ --- the familiar Bayesian MSE expression.
\item \textbf{Mutual information / log-posterior-ratio ($\Phi_{\text{MI}}$).}  $\Phi(\bx, \bs_{1:K};\, \bc) = \log \bb_K(\bx) - \log p_0(\bx)$.  Same argument: $\bs_{1:K}$ enters only through $\bb_K$.  Belief-encodable with $\Phi_0(\bx, \bb;\, \bc) = \log \bb(\bx) - \log p_0(\bx)$.  The belief-level average is $\E_{\bx \mid \bb}[\Phi_0] = \mathrm{KL}(\bb \Vert p_0)$; averaging further under the prior gives the mutual information $I(\bx;\, \by_{1:K})$.
\end{itemize}

The Fisher-information reward $\Phi_{\log J}(\bx, \bs_{1:K};\, \bc) = \tfrac{1}{2} \log \det J_N(\bx, \bs_{1:K};\, \bc)$ has an explicit $\bs_{1:K}$ dependence through the accumulated Fisher matrix $J_N$.  Two schedules that reach the same terminal posterior $\bb_K$ but use different action trajectories can have different $J_N$ values (the matrix accumulates a different sum) and therefore different $\Phi$. To address this formally, we can redefine the ``belief'' to carry whatever action-dependent accumulator the reward references: in the Fisher case, augment $\bb \to \tilde{\bb} = (\bb, J)$, so that $\Phi_{\log J}(\bx, \bs_{1:K};\, \bc)$ becomes a function of $(\bx, \tilde{\bb}_K, \bc)$.  The belief-encodability condition now holds on the augmented belief by construction.  The cost is a larger state space for the DP.

\bibliography{references}

\end{document}


\maketitle

\begin{abstract}
This tutorial is a self-contained pedagogical companion to the joint-DP framework of the main paper.  It builds the vocabulary of sequential Bayesian decision making from scratch --- states, beliefs, actions, policies, rewards --- states Bellman's principle of optimality in plain language and in formulas, and works the associated backward-induction recursion in full on two representative problems: the superconducting-qubit Ramsey flux sensor (continuous state, belief compressed to a count-tuple) and the radar beam-search POMDP (discrete state, belief lives natively on a probability simplex).  No prior exposure to reinforcement learning or Bellman equations is assumed; basic probability and elementary linear algebra suffice.
\end{abstract}

\tableofcontents

\section{What sequential decision making is, and why it needs its own theory}
\label{sec:intro}

A single Bayesian measurement is a one-shot decision: pick an experimental setting, record a datum, update a posterior, stop.  A \emph{sequential} measurement is a cascade of such decisions, with the crucial feature that each action is chosen \emph{after} seeing the outcomes of all previous ones.  This freedom to condition is what adaptive sensing trades on: a radar that scans slowly and widely is worth less than one that points at where the previous hit suggested a target might be; a quantum sensor that always uses the same Ramsey delay wastes information that a shorter first probe could have unlocked.

The mathematical awkwardness is that the optimal next action depends on whatever the posterior happens to look like when it is time to pick it --- and the posterior depends on every previous choice and every previous outcome, which are themselves random.  You cannot just ``optimize over action sequences'' the way you optimize over a single experimental setting, because an action sequence that performs well against one realization of the data performs poorly against another.  The right object to optimize is a \emph{policy}: a rule that maps whatever you currently believe to whatever you should do next.  Policies live in a much larger space than action sequences, and the infrastructure for finding the best one is dynamic programming --- the subject of this tutorial.

Two running examples carry the load.  The first is a superconducting-qubit Ramsey flux sensor: the unknown is a continuous scalar (the external flux), each action is a pair (Ramsey delay, repetition count), and each outcome is a Binomial count.  The second is a radar beam-search problem: the unknown is which of sixteen cells on a ring contains a target, each action is a beam pointing, and each outcome is a binary detection.  Both have a four-step measurement horizon.  The first forces us to confront a continuous-belief-space obstacle and solve it with a sufficient-statistic reduction; the second is natively finite and lets us see the DP in its rawest form.

\section{The building blocks}
\label{sec:blocks}

\subsection{The hidden state and the prior}

Nature draws a value $\bx$ from a known prior $p_0(\bx)$.  This value is fixed for the duration of the experiment but unknown to us.  In the qubit problem $\bx = \Phi_\mathrm{ext}$, a real number in some bounded interval; in the radar problem $\bx \in \{1, 2, \ldots, 16\}$, a discrete label for which cell the target is in.

\subsection{Actions, observations, and the likelihood}

At each of $K$ epochs we choose an action $\bs_k$ from an action set $\mathcal{S}$ and receive an observation $\by_k$ drawn from the likelihood
\begin{equation}
\by_k \sim p(\by \mid \bx, \bs_k;\, \bc),
\label{eq:likelihood}
\end{equation}
where $\bc$ is the physical hardware that parametrizes the likelihood.  In the qubit problem the action is $(\tau, n)$ --- a Ramsey delay and a repetition count --- and the observation is the number of ``1'' outcomes, distributed as Binomial$(n, p(\bx, \tau; \bc))$.  In the radar problem the action is a cell to point the beam at and the observation is a bit: $0$ for miss, $1$ for detection.

\subsection{The belief and its update}

After observing $k$ epochs of data we know more about $\bx$ than the prior specified.  Our state of knowledge is the posterior
\begin{equation}
\bb_k(\bx) \;=\; p(\bx \mid \bs_{1:k}, \by_{1:k};\, \bc),
\end{equation}
called the \emph{belief} at step $k$.  It is obtained from the prior by recursively applying Bayes' rule: $\bb_0 = p_0$, and after each new measurement
\begin{equation}
\bb_{k}(\bx) \;=\; \frac{p(\by_k \mid \bx, \bs_k;\, \bc)\, \bb_{k-1}(\bx)}{\int p(\by_k \mid \bx', \bs_k;\, \bc)\, \bb_{k-1}(\bx')\, d\bx'}.
\label{eq:bayes}
\end{equation}
We denote the mapping defined by Equation~\eqref{eq:bayes} by $T$: $\bb_k = T(\bb_{k-1}, \bs_k, \by_k;\, \bc)$.  Two facts about the belief that we will use repeatedly:

\begin{enumerate}[nosep, leftmargin=*]
\item The belief is a sufficient statistic of the history for inference about $\bx$.  Any expectation involving $\bx$ and conditioning on the history only needs $\bb_k$, not the raw $(\bs_{1:k}, \by_{1:k})$.
\item The belief is the natural state variable for the optimization problem.  A well-posed policy is a function of the current belief, $\bs_k = \pi(\bb_{k-1})$, rather than of the raw history.
\end{enumerate}

\subsection{The reward and the objective}

We need a scalar criterion to rank policies.  A pointwise reward $\Phi(\bx, \bs_{1:K};\, \bc)$ is a function that returns a number for each realization of the unknown state, the action sequence, and the hardware.  In the qubit problem we use the negative terminal posterior variance (the Bayesian MSE of the posterior-mean estimator is its prior average); in the radar problem we use the mutual information, equivalent to the negative terminal posterior entropy.

The \emph{value} of a policy $\pi$ is the expected reward when $\bx$ is drawn from the prior, actions are chosen by $\pi$, and observations are sampled from the likelihood:
\begin{equation}
V(\pi) \;=\; \E_{\bx \sim p_0}\, \E_{\by_{1:K} \mid \bx, \pi,\, \bc}\!\bigl[\Phi(\bx, \bs_{1:K}^\pi;\, \bc)\bigr].
\label{eq:V_pi}
\end{equation}
The superscript on $\bs^\pi$ is a reminder that the action sequence is not fixed --- each $\bs_k$ is produced by applying $\pi$ to the current belief, which depends on the random trajectory $(\by_{1:k-1})$.

The \emph{optimal policy} is any $\pi^\star$ that maximizes $V(\pi)$.  Solving for $\pi^\star$ is the central task of sequential decision theory.

\subsection{Why we cannot just optimize over action sequences}

If the horizon were $K=1$ the optimal policy would reduce to a single best action $\bs^\star = \argmax_{\bs} \E_{\bx, \by}[\Phi(\bx, \bs; \bc)]$ --- ordinary Bayesian optimal experimental design.  For $K \ge 2$ two obstructions appear.  First, the optimal $\bs_2$ depends on $\by_1$, which is random.  So there is no single ``best $\bs_2$''; the best choice is a \emph{function} of $\by_1$.  Second, the optimal $\bs_1$ should anticipate how its outcome will shape the best $\bs_2$ --- in other words, $\bs_1$ must account for the option value of information.  Optimizing each epoch separately in a greedy or myopic fashion misses both of these couplings.

The language that lets us talk about this coupling is Bellman's.

\section{Bellman's principle of optimality}
\label{sec:bellman_principle}

\subsection{The principle in plain language}

An optimal policy has the following property.  If you follow it up to some moment, reach whatever belief you happen to reach, and then ask ``what is the best I can still do from here?'' --- the continuation of the optimal policy from that belief onward is itself an optimal policy for the sub-problem starting at that belief.  Nothing the first part of the policy did will have rendered the second part suboptimal.

That is almost circular, but the content is sharp: the globally optimal policy is \emph{recursive}.  You can assemble it by solving the sub-problems (one step before the end, then two before, then three, and so on) and gluing their optimal pieces together.

\subsection{The principle in formulas}

Fix the hardware $\bc$.  Define the \emph{value-to-go} function
\begin{equation}
V_k(\bb) \;=\; \max_{\pi}\, \E\bigl[\Phi_0(\bx, \bb_K;\, \bc) \,\big|\, \bb_k = \bb,\ \text{policy $\pi$ from step $k$ onward}\bigr],
\label{eq:V_k_def}
\end{equation}
where $\Phi_0(\bx, \bb_K; \bc)$ is the reward written as a function of the unknown state and the terminal belief (for our two examples $\Phi_0$ is $-(\bx - \hat{x})^2$ for the qubit, or $\log \bb_K(\bx) - \log p_0(\bx)$ for the radar; see Appendix~\ref{app:phi0}).

$V_k(\bb)$ is the best expected score an optimal agent can still achieve if the current belief is $\bb$ and there are $K - k$ measurements still to take.  Bellman's principle says that this function satisfies the recursion
\begin{align}
V_K(\bb) &\;=\; \E_{\bx \mid \bb}\!\bigl[\Phi_0(\bx, \bb;\, \bc)\bigr], \label{eq:bellman_terminal}\\
V_k(\bb) &\;=\; \max_{\bs \in \mathcal{S}}\; \E_{\by \mid \bb, \bs,\, \bc}\!\bigl[V_{k+1}\bigl(T(\bb, \bs, \by;\, \bc)\bigr)\bigr], \quad k = K-1, \ldots, 0.
\label{eq:bellman_general}
\end{align}
Equation~\eqref{eq:bellman_terminal} is the \emph{terminal condition}: at step $K$ the experiment is over and the value-to-go collapses to the reward evaluated at whatever belief we have arrived at.  Equation~\eqref{eq:bellman_general} is the \emph{Bellman backup}: to know the value-to-go at belief $\bb$ with $K - k$ steps left, try every action, for each one compute the expected continuation value (which is $V_{k+1}$ evaluated at the next belief, averaged over the observation that the action would produce), and take the best.  The argmax is the optimal action at belief $\bb$ with $K - k$ steps left:
\begin{equation}
\pi^\star_k(\bb) \;=\; \argmax_{\bs \in \mathcal{S}}\; \E_{\by \mid \bb, \bs,\, \bc}\!\bigl[V_{k+1}\bigl(T(\bb, \bs, \by;\, \bc)\bigr)\bigr].
\label{eq:pi_star}
\end{equation}
Reading $\pi^\star_0, \pi^\star_1, \ldots, \pi^\star_{K-1}$ off the recursion gives the globally optimal adaptive policy.

\subsection{Deriving the recursion from \texorpdfstring{$V(\pi)$}{V(pi)}}
\label{sec:derivation}

We stated the recursion above.  Now we derive it --- starting from Equation~\eqref{eq:V_pi}, the expected reward under a fixed policy --- in four steps.  Each step isolates one ingredient: change of conditioning, sufficiency of the posterior, trajectory factorization through the belief update, and Bellman's principle applied as an interchange of max and expectation.

\paragraph{Step 1: Condition on the data, not on the state.}  Equation~\eqref{eq:V_pi} is a nested expectation, outer over the prior $p_0$ and inner over the data under $\pi$.  By the law of total expectation (Fubini plus the tower rule), we may swap these:
\begin{equation}
V(\pi) \;=\; \E_{\by_{1:K} \mid \pi,\, \bc}\!\bigl[\, \E_{\bx \mid \by_{1:K},\, \bc}\!\bigl[\Phi(\bx, \bs_{1:K}^\pi;\, \bc)\bigr] \,\bigr].
\label{eq:V_pi_swap}
\end{equation}
The outer expectation is now over how the data trajectory plays out under $\pi$; the inner expectation, over what $\bx$ could have been given the observed data.

\paragraph{Step 2: Replace the history-dependent integrand by a belief-dependent one.}  Two logically separate substitutions happen here, both often compressed into the single phrase ``sufficient-statistic reduction.''  They are worth separating, because the first is automatic and the second is a real assumption on the reward.

\textbf{(2a) Conditioning change: $p(\bx \mid \by_{1:K}) = \bb_K$.}  The posterior $\bb_K$ is by definition the conditional distribution of $\bx$ given the measurement history:
\begin{equation}
\bb_K(\bx) \;=\; p(\bx \mid \by_{1:K}, \bs_{1:K};\, \bc).
\end{equation}
Consequently, for any integrand $f(\bx)$ --- no matter how chosen ---
\begin{equation}
\E_{\bx \mid \by_{1:K},\, \bc}\!\bigl[f(\bx)\bigr] \;=\; \int f(\bx)\, \bb_K(\bx)\, d\bx \;=\; \E_{\bx \mid \bb_K}\!\bigl[f(\bx)\bigr].
\label{eq:suff_of_bK}
\end{equation}
This is just the definition of conditional expectation.  The ``history'' $\by_{1:K}$ and the ``belief'' $\bb_K$ are interchangeable as the conditioning object for any function of $\bx$.  Point (2a) is therefore free.

\textbf{(2b) Function substitution: $\Phi(\bx, \bs_{1:K};\, \bc) = \Phi_0(\bx, \bb_K;\, \bc)$.}  This is the part that requires care.  The original reward $\Phi$ takes the action trajectory $\bs_{1:K}$ as an argument; we want to replace that argument by the terminal belief $\bb_K$.  The substitution is legitimate iff $\Phi$'s dependence on $\bs_{1:K}$ actually factors through $\bb_K$.

\emph{Definition (belief-encodable reward).}  We call $\Phi$ \emph{belief-encodable} if there is a function $\Phi_0(\bx, \bb;\, \bc)$ such that, for every $\bx$, every $\bc$, and every pair of trajectories $(\bs_{1:K}, \by_{1:K})$ and $(\bs'_{1:K}, \by'_{1:K})$ whose terminal beliefs agree ($\bb_K = \bb'_K$),
\begin{equation}
\Phi(\bx, \bs_{1:K};\, \bc) \;=\; \Phi(\bx, \bs'_{1:K};\, \bc) \;=\; \Phi_0(\bx, \bb_K;\, \bc).
\label{eq:belief_encodable}
\end{equation}
In words: any two histories that land on the same terminal posterior yield the same reward.  If Equation~\eqref{eq:belief_encodable} holds, $\Phi_0$ is uniquely defined (up to values on unreachable beliefs), and substituting the rewrite into Equation~\eqref{eq:V_pi_swap} is valid.

\emph{Two easy cases.}  The two rewards used in this tutorial both pass the test immediately, because each is a simple functional of the posterior's shape:

\begin{itemize}[nosep, leftmargin=*]
\item \textbf{Posterior variance ($\Phi_{\text{Var}}$).}  $\Phi(\bx, \bs_{1:K};\, \bc) = -(\bx - \hat{x}(\bb_K))^2$, where $\hat{x}(\bb_K) = \E[\bx \mid \bb_K]$ is the posterior mean.  The only place $\bs_{1:K}$ enters is through $\bb_K$, and $\bb_K$ is itself the object we are substituting in.  Belief-encodable with $\Phi_0(\bx, \bb;\, \bc) = -(\bx - \hat{x}(\bb))^2$.  The belief-level average $\E_{\bx \mid \bb}[\Phi_0]$ equals $-\mathrm{Var}(\bx \mid \bb)$ --- the familiar Bayesian MSE expression.
\item \textbf{Mutual information / log-posterior-ratio ($\Phi_{\text{MI}}$).}  $\Phi(\bx, \bs_{1:K};\, \bc) = \log \bb_K(\bx) - \log p_0(\bx)$.  Same argument: $\bs_{1:K}$ enters only through $\bb_K$.  Belief-encodable with $\Phi_0(\bx, \bb;\, \bc) = \log \bb(\bx) - \log p_0(\bx)$.  The belief-level average is $\E_{\bx \mid \bb}[\Phi_0] = \mathrm{KL}(\bb \Vert p_0)$; averaging further under the prior gives the mutual information $I(\bx;\, \by_{1:K})$.
\end{itemize}

\emph{A case where the test fails.}  The Fisher-information-aligned reward $\Phi_{\log J}(\bx, \bs_{1:K};\, \bc) = \tfrac{1}{2} \log \det J_N(\bx, \bs_{1:K};\, \bc)$ has an explicit $\bs_{1:K}$ dependence through the accumulated Fisher matrix $J_N$.  Two schedules that reach the same terminal posterior $\bb_K$ but use different delays can have different $J_N$ values (the matrix accumulates a different sum) and therefore different $\Phi$.  Equation~\eqref{eq:belief_encodable} fails: same $\bb_K$, different $\Phi$.  $\Phi_{\log J}$ is \emph{not} belief-encodable as stated.

\emph{The repair.}  When Equation~\eqref{eq:belief_encodable} fails, redefine the ``belief'' to carry whatever action-dependent accumulator the reward references --- in the Fisher case, augment $\bb \to \tilde{\bb} = (\bb, J)$, so that $\Phi_{\log J}(\bx, \bs_{1:K};\, \bc)$ becomes a function of $(\bx, \tilde{\bb}_K, \bc)$.  The belief-encodability condition now holds on the augmented belief by construction.  The cost is a larger state space for the DP; the benefit is that Step~2b works again.  See §\ref{sec:augmentation} for the full treatment.

\textbf{Combining (2a) and (2b).}  Whenever the reward is belief-encodable (possibly after augmentation),
\begin{equation}
\E_{\bx \mid \by_{1:K},\, \bc}\!\bigl[\Phi(\bx, \bs_{1:K}^\pi;\, \bc)\bigr]
\;\stackrel{\text{(2b)}}{=}\;
\E_{\bx \mid \by_{1:K},\, \bc}\!\bigl[\Phi_0(\bx, \bb_K;\, \bc)\bigr]
\;\stackrel{\text{(2a)}}{=}\;
\E_{\bx \mid \bb_K}\!\bigl[\Phi_0(\bx, \bb_K;\, \bc)\bigr].
\end{equation}
The first step replaces the integrand by its belief-indexed form (2b); the second step changes the conditioning object (2a), which is automatic.  Plugging into Equation~\eqref{eq:V_pi_swap} gives
\begin{equation}
V(\pi) \;=\; \E_{\by_{1:K} \mid \pi,\, \bc}\!\bigl[\, \E_{\bx \mid \bb_K}\!\bigl[\Phi_0(\bx, \bb_K;\, \bc)\bigr] \,\bigr].
\label{eq:V_pi_belief}
\end{equation}
The inner bracket is a pure functional of $\bb_K$; write $\Phi_0^{\mathrm{belief}}(\bb;\, \bc) \coloneqq \E_{\bx \mid \bb}[\Phi_0(\bx, \bb;\, \bc)]$ to give it a name.  The two examples above illustrate the two most common forms of $\Phi_0^{\mathrm{belief}}$: the posterior variance and the posterior KL to the prior.

\paragraph{Step 3: Factorize the trajectory expectation through the belief update.}  Any realized trajectory $(\by_1, \ldots, \by_K)$ can be sampled one step at a time.  Draw $\by_1$ from $p(\cdot \mid \bb_0, \pi(\bb_0);\, \bc)$, update the belief via $\bb_1 = T(\bb_0, \pi(\bb_0), \by_1;\, \bc)$, draw $\by_2$ from $p(\cdot \mid \bb_1, \pi(\bb_1);\, \bc)$, and so on.  The outer expectation in Equation~\eqref{eq:V_pi_belief} thus factorizes as
\begin{equation}
V(\pi) \;=\; \E_{\by_1 \mid \bb_0, \pi, \bc}\, \E_{\by_2 \mid \bb_1, \pi, \bc}\, \cdots\, \E_{\by_K \mid \bb_{K-1}, \pi, \bc}\!\bigl[\,\Phi_0^{\mathrm{belief}}(\bb_K;\, \bc)\,\bigr].
\label{eq:V_factored}
\end{equation}

Define the \emph{policy-evaluation} value-to-go under a fixed policy $\pi$ as
\begin{equation}
U^\pi_k(\bb;\, \bc) \;\coloneqq\; \E_{\by_{k+1}, \ldots, \by_K \mid \bb_k = \bb, \pi, \bc}\!\bigl[\Phi_0^{\mathrm{belief}}(\bb_K;\, \bc)\bigr].
\label{eq:U_def}
\end{equation}
Equation~\eqref{eq:V_factored} is exactly $V(\pi) = U^\pi_0(p_0;\, \bc)$.  The terminal case of Equation~\eqref{eq:U_def} is immediate: with no remaining measurements, the expectation degenerates and
\begin{equation}
U^\pi_K(\bb;\, \bc) \;=\; \Phi_0^{\mathrm{belief}}(\bb;\, \bc).
\label{eq:U_terminal}
\end{equation}

\textbf{Deriving the one-step recursion.}  We now show how $U^\pi_{k-1}$ is determined by $U^\pi_k$, which is the content of the backward recursion.  Start from the definition, Equation~\eqref{eq:U_def}, at step $k - 1$:
\begin{equation}
U^\pi_{k-1}(\bb;\, \bc) \;=\; \E_{\by_k, \by_{k+1}, \ldots, \by_K \mid \bb_{k-1} = \bb, \pi, \bc}\!\bigl[\Phi_0^{\mathrm{belief}}(\bb_K;\, \bc)\bigr].
\label{eq:U_km1}
\end{equation}
Apply the tower rule to split the expectation over $\by_k$ from the expectation over the remaining trajectory:
\begin{equation}
U^\pi_{k-1}(\bb;\, \bc) \;=\; \E_{\by_k \mid \bb_{k-1} = \bb, \pi(\bb), \bc}\!\Biggl[\,\underbrace{\E_{\by_{k+1}, \ldots, \by_K \mid \bb_{k-1} = \bb, \by_k, \pi, \bc}\!\bigl[\Phi_0^{\mathrm{belief}}(\bb_K;\, \bc)\bigr]}_{\text{inner expectation}}\,\Biggr].
\label{eq:U_km1_tower}
\end{equation}
Now examine the inner expectation.  Its conditioning variables are $(\bb_{k-1} = \bb,\ \by_k,\ \pi,\ \bc)$.  Given these, the belief at step $k$ is determined:
\begin{equation}
\bb_k \;=\; T\bigl(\bb, \pi(\bb), \by_k;\, \bc\bigr).
\label{eq:bk_det}
\end{equation}
The distribution of the remaining trajectory $(\by_{k+1}, \ldots, \by_K)$ depends on the past only through the current belief (the \emph{sufficiency of the belief} for the future dynamics, in the Markov-chain sense).  So conditioning on $(\bb_{k-1} = \bb,\ \by_k)$ is equivalent --- for the purpose of computing expectations of functions of the future --- to conditioning on the single number $\bb_k$ from Equation~\eqref{eq:bk_det}:
\begin{equation}
\E_{\by_{k+1}, \ldots, \by_K \mid \bb_{k-1} = \bb, \by_k, \pi, \bc}\!\bigl[\,\cdot\,\bigr]
\;=\;
\E_{\by_{k+1}, \ldots, \by_K \mid \bb_k = T(\bb, \pi(\bb), \by_k;\, \bc), \pi, \bc}\!\bigl[\,\cdot\,\bigr].
\label{eq:belief_sufficient}
\end{equation}
The right-hand side of Equation~\eqref{eq:belief_sufficient}, applied to the integrand $\Phi_0^{\mathrm{belief}}(\bb_K;\, \bc)$, is exactly the definition of $U^\pi_k$, Equation~\eqref{eq:U_def}, evaluated at belief $T(\bb, \pi(\bb), \by_k;\, \bc)$:
\begin{equation}
\E_{\by_{k+1}, \ldots, \by_K \mid \bb_k = T(\ldots), \pi, \bc}\!\bigl[\Phi_0^{\mathrm{belief}}(\bb_K;\, \bc)\bigr]
\;=\; U^\pi_k\bigl(T(\bb, \pi(\bb), \by_k;\, \bc);\, \bc\bigr).
\label{eq:inner_is_Uk}
\end{equation}
Substituting Equation~\eqref{eq:inner_is_Uk} back into Equation~\eqref{eq:U_km1_tower}:
\begin{equation}
\boxed{\ U^\pi_{k-1}(\bb;\, \bc) \;=\; \E_{\by_k \mid \bb, \pi(\bb), \bc}\!\bigl[\,U^\pi_k\bigl(T(\bb, \pi(\bb), \by_k;\, \bc);\, \bc\bigr)\,\bigr]\ }
\label{eq:U_recursion}
\end{equation}
which is the policy-evaluation recursion.  Shifting the index by one ($k - 1 \to k,\ k \to k + 1$) and pairing with Equation~\eqref{eq:U_terminal} gives the full backward recursion for $U^\pi$.

\textbf{What each step used.}  Three ingredients, one per substitution:
\begin{enumerate}[nosep, leftmargin=*]
\item Equation~\eqref{eq:U_km1_tower} uses the \emph{tower rule} (iterated expectation): joint expectations factor into nested conditional expectations.  Applied to separate ``first step'' from ``rest.''
\item Equation~\eqref{eq:bk_det} uses the \emph{deterministic belief update} $T$: given the previous belief, action, and outcome, the next belief is fixed by Bayes' rule with no remaining randomness.
\item Equation~\eqref{eq:belief_sufficient} uses the \emph{Markov property of the belief}: the future trajectory's distribution depends on the past only through the current belief.  Every piece of history that is not captured in $\bb_k$ drops out.
\end{enumerate}
No optimization is used anywhere in this derivation; Equation~\eqref{eq:U_recursion} is a valid identity for every fixed policy $\pi$.  The backward recursion for $U^\pi$ is therefore the right-hand object to carry into Step~4, where we maximize over $\pi$ to turn it into the Bellman recursion for $V_k$.

\paragraph{Step 4: Maximize over $\pi$ (Bellman's principle).}  We now take $\max$ over $\pi$ of the policy-evaluation recursion of Step 3 to get the Bellman recursion.  The step has eight substeps; we go through each one.

\subparagraph{Step 4a --- Definition.}  Define
\begin{equation}
V_k(\bb;\, \bc) \;\coloneqq\; \max_\pi\, U^\pi_k(\bb;\, \bc),
\label{eq:V_k_from_U}
\end{equation}
the value-to-go under the \emph{optimal} continuation from belief $\bb$ at step $k$.  Our goal is a recursion for $V_k$ in terms of $V_{k+1}$.

\subparagraph{Step 4b --- What goes wrong if we apply $\max_\pi$ naively.}  The policy-evaluation recursion Equation~\eqref{eq:U_recursion} reads
\begin{equation}
U^\pi_k(\bb;\, \bc) \;=\; \E_{\by_{k+1} \mid \bb, \pi(\bb),\, \bc}\!\bigl[\,U^\pi_{k+1}\bigl(T(\bb, \pi(\bb), \by_{k+1};\, \bc);\, \bc\bigr)\,\bigr].
\end{equation}
If we try to take $\max_\pi$ of both sides and naively push it through $\E_{\by_{k+1}}$, we would need
\begin{equation}
\max_\pi\, \int f(\pi, \by)\, p(\by)\, d\by \;\stackrel{?}{=}\; \int \max_\pi\, \bigl[f(\pi, \by)\, p(\by)\bigr]\, d\by.
\end{equation}
In general these are not equal --- the $\max$ and the integral do not commute for an arbitrary family of functions indexed by $\pi$.  So we have to argue why they commute in our setting.  The argument has two ingredients: the \emph{decomposability} of $\pi$, and the \emph{causality} of the measurement.

\subparagraph{Step 4c --- Decompose $\pi$.}  At step $k$, a policy $\pi$ determines two pieces of information:
\begin{itemize}[nosep, leftmargin=*]
\item The action at the current belief, $\bs_k = \pi(\bb)$ --- a single value in the action set.
\item The continuation $\pi'$ from step $k+1$ onward --- a function that maps future beliefs to future actions at steps $k+1, k+2, \ldots, K-1$.
\end{itemize}
Write this as $\pi = (\bs_k, \pi')$.  Substituting into the policy-evaluation recursion,
\begin{equation}
U^\pi_k(\bb;\, \bc) \;=\; \E_{\by_{k+1} \mid \bb, \bs_k,\, \bc}\!\bigl[\, U^{\pi'}_{k+1}\bigl(T(\bb, \bs_k, \by_{k+1};\, \bc);\, \bc\bigr)\,\bigr],
\end{equation}
where the right-hand side now depends on $\pi$ only through the pair $(\bs_k, \pi')$.  Maximizing over $\pi$ is therefore equivalent to maximizing jointly over $(\bs_k, \pi')$:
\begin{equation}
V_k(\bb;\, \bc) \;=\; \max_{\bs_k \in \mathcal{S}}\,\max_{\pi'}\; \E_{\by_{k+1} \mid \bb, \bs_k,\, \bc}\!\bigl[\, U^{\pi'}_{k+1}\bigl(T(\bb, \bs_k, \by_{k+1};\, \bc);\, \bc\bigr)\,\bigr].
\label{eq:double_max_tutorial}
\end{equation}
Fix $\bs_k$ for the remainder of Steps 4d--4f; we will return to the outer $\max_{\bs_k}$ at the end.

\subparagraph{Step 4d --- Two observations about $(y_{k+1}, \pi')$.}  Write the one-step expectation out as an integral:
\begin{equation}
\max_{\pi'} \int U^{\pi'}_{k+1}\bigl(T(\bb, \bs_k, \by_{k+1};\, \bc);\, \bc\bigr)\, p(\by_{k+1} \mid \bb, \bs_k, \bc)\, d\by_{k+1}.
\label{eq:max_integral}
\end{equation}
Two structural facts about this expression are what make the interchange go through.

\textbf{Observation 1 (causality): $p(\by_{k+1} \mid \bb, \bs_k, \bc)$ does not depend on $\pi'$.}  The measurement $\by_{k+1}$ is drawn from the likelihood $p(\cdot \mid \bb, \bs_k, \bc)$, which depends only on the current belief $\bb$ and the current action $\bs_k$ --- not on the continuation $\pi'$.  Causally, $\by_{k+1}$ is generated \emph{before} $\pi'$ is applied; the continuation policy only enters the picture once $T(\bb, \bs_k, \by_{k+1};\, \bc)$ has been computed.  So $p(\by_{k+1} \mid \bb, \bs_k, \bc)$ is a $\pi'$-free non-negative weight.

\textbf{Observation 2 ($\pi'$ is a function, not a value): $\pi'$ spans a function space.}  A continuation policy takes a belief and returns an action, $\pi' : \mathcal{B} \to \mathcal{S}$.  The belief at which $\pi'$ is evaluated, $T(\bb, \bs_k, \by_{k+1};\, \bc)$, depends on $\by_{k+1}$, so different realizations of $\by_{k+1}$ cause $\pi'$ to be evaluated at different beliefs, where it is allowed to take different action values.  The $\max$ over $\pi'$ is therefore a $\max$ over a \emph{function space}, not over a single action value.

\paragraph{The interchange lemma.}  Before completing Step 4e, we isolate the abstract identity that does the work in a stand-alone lemma.  Step 4e will then be a one-line application.

\begin{framed}
\noindent\textbf{Lemma (max--integral interchange over a function space).}  Let $Y$ be a measure space with non-negative measure $\mu$, $A$ a set (the ``action space''), and $g: Y \times A \to \R$ a measurable function.  Let $\mathcal{C}$ be a class of admissible measurable functions $\phi: Y \to A$ such that $\int g(y, \phi(y))\, d\mu(y)$ is well-defined for every $\phi \in \mathcal{C}$.  Then
\begin{equation}
\max_{\phi \in \mathcal{C}}\, \int g(y, \phi(y))\, d\mu(y) \;=\; \int \max_{a \in A}\, g(y, a)\, d\mu(y)
\label{eq:interchange_lemma}
\end{equation}
if and only if
\begin{enumerate}[nosep, label=(\roman*)]
\item \textbf{(attainment)} for each $y \in Y$, the pointwise argmax $a^\star(y) \coloneqq \argmax_{a \in A}\, g(y, a)$ exists, and
\item \textbf{(admissibility)} the assembled function $a^\star(\cdot)$ lies in $\mathcal{C}$.
\end{enumerate}

\vspace{2pt}
\textbf{Proof.}  One inequality always holds; the other requires (i) and (ii).

\emph{Step A (LHS $\leq$ RHS, unconditional).}  For any admissible $\phi \in \mathcal{C}$ and any $y \in Y$,
\begin{equation*}
g(y, \phi(y)) \;\leq\; \max_{a \in A}\, g(y, a),
\end{equation*}
by definition of the pointwise maximum.  Integrating both sides against $\mu \geq 0$ preserves the inequality,
\begin{equation*}
\int g(y, \phi(y))\, d\mu(y) \;\leq\; \int \max_{a \in A}\, g(y, a)\, d\mu(y).
\end{equation*}
The right-hand side does not depend on $\phi$, so it bounds the max over $\phi$ as well:
\begin{equation*}
\max_{\phi \in \mathcal{C}}\, \int g(y, \phi(y))\, d\mu(y) \;\leq\; \int \max_{a \in A}\, g(y, a)\, d\mu(y).
\end{equation*}

\emph{Step B (LHS $\geq$ RHS under (i) and (ii)).}  Define $\phi^\star \coloneqq a^\star$.  By (i), $\phi^\star(y)$ is well-defined at every $y$; by (ii), $\phi^\star \in \mathcal{C}$.  At every $y$,
\begin{equation*}
g(y, \phi^\star(y)) \;=\; g(y, a^\star(y)) \;=\; \max_{a \in A}\, g(y, a),
\end{equation*}
with the first equality by definition of $\phi^\star$ and the second by definition of $a^\star$ via (i).  Integrating,
\begin{equation*}
\int g(y, \phi^\star(y))\, d\mu(y) \;=\; \int \max_{a \in A}\, g(y, a)\, d\mu(y).
\end{equation*}
Since $\phi^\star \in \mathcal{C}$, the max over $\mathcal{C}$ is at least this value:
\begin{align*}
\max_{\phi \in \mathcal{C}}\, \int g(y, \phi(y))\, d\mu(y)
&\;\geq\; \int g(y, \phi^\star(y))\, d\mu(y) \\
&\;=\; \int \max_{a \in A}\, g(y, a)\, d\mu(y).
\end{align*}

\emph{Step C (combine).}  Step A gives LHS $\leq$ RHS; Step B gives LHS $\geq$ RHS.  Together, LHS $=$ RHS, which is Equation~\eqref{eq:interchange_lemma}.

\emph{Step D (strict inequality when (i) or (ii) fails).}  If (i) fails on a $\mu$-positive set of $y$'s, the RHS integrand uses a $\sup$ rather than an attained $\max$ at those $y$'s, and no $\phi \in \mathcal{C}$ can match it everywhere; LHS is strictly less than RHS by the sup gap.  If (ii) fails --- $a^\star$ exists pointwise but is not in $\mathcal{C}$ --- then no $\phi \in \mathcal{C}$ achieves the pointwise max at every $y$, and again LHS is strictly less than RHS.  $\square$

\vspace{4pt}
\textbf{Operational reading of the conditions.}
\begin{itemize}[nosep, leftmargin=*]
\item \textbf{(i) Attainment} says the inner max over actions is \emph{achieved} (not just approximated) at each $y$.  Holds automatically when $A$ is finite or when $A$ is compact and $g(y, \cdot)$ is upper semicontinuous.
\item \textbf{(ii) Admissibility} says the pointwise-argmax function, which in general depends on $y$, must itself be an admissible element of $\mathcal{C}$.  Equivalently, $\mathcal{C}$ must be a function class rich enough to include all such $y$-dependent choices.  This is satisfied when $\mathcal{C}$ is the class of all measurable $\phi: Y \to A$, and it fails if $\mathcal{C}$ is restricted --- e.g., to constant functions (``fixed schedules''), in which case LHS $<$ RHS strictly.
\end{itemize}
\end{framed}

\subparagraph{Step 4e --- The interchange (load-bearing step).}  Apply the lemma with
\begin{itemize}[nosep, leftmargin=*]
\item $Y$ = the space of possible values of $\by_{k+1}$;
\item measure $d\mu(\by_{k+1}) = p(\by_{k+1} \mid \bb, \bs_k, \bc)\, d\by_{k+1}$ (the outcome distribution, non-negative and $\pi'$-free by Observation 1);
\item $A$ = the space of admissible continuation policies from step $k+1$ onward;
\item $\mathcal{C}$ = the class of all measurable continuation policies (equivalently, history-measurable functions from step-$k+1$ beliefs to actions through step $K-1$);
\item $g(\by_{k+1}, \pi') = U^{\pi'}_{k+1}\bigl(T(\bb, \bs_k, \by_{k+1};\, \bc);\, \bc\bigr)$.
\end{itemize}
Both conditions of the lemma are satisfied.  \emph{(i) Attainment}: for each $\by_{k+1}$, the inner problem $\max_{\pi'} U^{\pi'}_{k+1}(T(\cdots);\, \bc)$ is exactly the Bellman sub-problem at step $k+1$, which is attained whenever the inner action set is finite (as in all case studies in this tutorial) or compact with continuous $U^{\pi'}_{k+1}$.  \emph{(ii) Admissibility}: the pointwise argmax
\begin{equation*}
\pi^\star(\by_{k+1}) \;\coloneqq\; \argmax_{\pi'}\, U^{\pi'}_{k+1}\bigl(T(\bb, \bs_k, \by_{k+1};\, \bc);\, \bc\bigr)
\end{equation*}
is a $\by_{k+1}$-dependent continuation, which Observation 2 tells us is admissible (adaptive continuations can depend on the realized $\by_{k+1}$).

The lemma therefore gives
\begin{align}
&\max_{\pi'}\, \int U^{\pi'}_{k+1}\bigl(T(\bb, \bs_k, \by_{k+1};\, \bc);\, \bc\bigr)\, p(\by_{k+1} \mid \bb, \bs_k, \bc)\, d\by_{k+1} \notag \\
&\qquad\;=\; \int \max_{\pi'}\, U^{\pi'}_{k+1}\bigl(T(\bb, \bs_k, \by_{k+1};\, \bc);\, \bc\bigr)\, p(\by_{k+1} \mid \bb, \bs_k, \bc)\, d\by_{k+1}.
\label{eq:interchange}
\end{align}
(The weight $p(\by_{k+1} \mid \bb, \bs_k, \bc)$ on the right is pulled out of the $\max_{\pi'}$ because it is a non-negative $\pi'$-free constant, exactly as Observation 1 allows.  The lemma does the interchange; this factor-out is the easy follow-on step.)  If the admissibility condition failed --- if $\pi'$ were restricted to a non-adaptive class such as constant schedules --- the lemma's condition (ii) would be violated and the interchange would produce strict inequality.

\subparagraph{Step 4f --- Pull $p$ out of the inner $\max$ (easy step).}  By Observation 1, $p(\by_{k+1} \mid \bb, \bs_k, \bc)$ is a $\pi'$-free non-negative constant as far as the inner $\max_{\pi'}$ is concerned.  Non-negative constants commute with $\max$:
\begin{equation}
\max_{\pi'}\, \bigl[\, U^{\pi'}_{k+1}\, p(\by_{k+1} \mid \bb, \bs_k, \bc)\, \bigr] \;=\; p(\by_{k+1} \mid \bb, \bs_k, \bc)\, \max_{\pi'}\, U^{\pi'}_{k+1}.
\label{eq:factor_out}
\end{equation}
Substituting Equation~\eqref{eq:factor_out} into Equation~\eqref{eq:interchange}:
\begin{align}
\max_{\pi'}\, \E_{\by_{k+1} \mid \bb, \bs_k, \bc}\!\bigl[\, U^{\pi'}_{k+1}\, \bigr]
\;&=\; \int p(\by_{k+1} \mid \bb, \bs_k, \bc)\, \max_{\pi'}\, U^{\pi'}_{k+1}\bigl(T(\bb, \bs_k, \by_{k+1};\, \bc);\, \bc\bigr)\, d\by_{k+1} \notag \\
\;&=\; \E_{\by_{k+1} \mid \bb, \bs_k, \bc}\!\bigl[\, \max_{\pi'}\, U^{\pi'}_{k+1}\bigl(T(\bb, \bs_k, \by_{k+1};\, \bc);\, \bc\bigr)\, \bigr].
\label{eq:max_inside_E}
\end{align}

\subparagraph{Step 4g --- Substitute the definition of $V_{k+1}$.}  By Equation~\eqref{eq:V_k_from_U} applied at step $k+1$,
\begin{equation}
\max_{\pi'}\, U^{\pi'}_{k+1}(\bb';\, \bc) \;=\; V_{k+1}(\bb';\, \bc)\qquad \text{for any belief } \bb'.
\end{equation}
Applied at $\bb' = T(\bb, \bs_k, \by_{k+1};\, \bc)$ inside the expectation of Equation~\eqref{eq:max_inside_E}:
\begin{equation}
\max_{\pi'}\, U^{\pi'}_{k+1}\bigl(T(\bb, \bs_k, \by_{k+1};\, \bc);\, \bc\bigr) \;=\; V_{k+1}\bigl(T(\bb, \bs_k, \by_{k+1};\, \bc);\, \bc\bigr).
\end{equation}
So Equation~\eqref{eq:max_inside_E} becomes
\begin{equation}
\max_{\pi'}\, \E_{\by_{k+1} \mid \bb, \bs_k, \bc}\!\bigl[\, U^{\pi'}_{k+1}\, \bigr]
\;=\; \E_{\by_{k+1} \mid \bb, \bs_k, \bc}\!\bigl[\, V_{k+1}\bigl(T(\bb, \bs_k, \by_{k+1};\, \bc);\, \bc\bigr)\, \bigr].
\end{equation}

\subparagraph{Step 4h --- Put the outer $\max_{\bs_k}$ back and finish.}  Substituting into Equation~\eqref{eq:double_max_tutorial}:
\begin{equation}
V_k(\bb;\, \bc) \;=\; \max_{\bs_k \in \mathcal{S}}\; \E_{\by_{k+1} \mid \bb, \bs_k, \bc}\!\bigl[\, V_{k+1}\bigl(T(\bb, \bs_k, \by_{k+1};\, \bc);\, \bc\bigr)\, \bigr],
\end{equation}
which is exactly the Bellman recursion, Equation~\eqref{eq:bellman_general}.  The terminal condition Equation~\eqref{eq:bellman_terminal} is immediate from Equation~\eqref{eq:U_terminal}: at step $K$ there is no further action to take, $U^\pi_K(\bb;\, \bc)$ does not depend on $\pi$, so $V_K(\bb;\, \bc) = \max_\pi U^\pi_K(\bb;\, \bc) = U^\pi_K(\bb;\, \bc) = \E_{\bx \mid \bb}[\Phi_0(\bx, \bb;\, \bc)]$.

Evaluating the Bellman recursion at $k = 0$, $\bb = p_0$ recovers $V^\star = \max_\pi V(\pi) = V_0(p_0;\, \bc)$, closing the loop from Equation~\eqref{eq:V_pi} to Equations~\eqref{eq:bellman_terminal}--\eqref{eq:bellman_general}.

\paragraph{Where the adaptive-versus-fixed distinction lives.}  The only non-trivial move in the derivation is Step 4e, the interchange of $\max_{\pi'}$ with $\E_{\by_{k+1}}$.  That interchange is valid \emph{because} $\pi'$ is admissible as a $\by_{k+1}$-measurable function (Observation 2), so the pointwise argmax over $\by_{k+1}$ is a legitimate element of the class we are maximizing over.  If we restricted $\pi'$ to a fixed schedule of actions (non-adaptive), Observation 2 would fail, $\max_{\pi'}$ would \emph{not} commute with $\E_{\by_{k+1}}$, and the Bellman recursion would not simplify.  What one would recover instead is the \emph{fixed-schedule} value $V_{\text{fixed}}$, strictly smaller than $V_{\text{adaptive}}$ in general.  The gap between the two --- the ``adaptive gain'' that the companion paper's EVPI analysis quantifies --- is in a precise sense the value of Step 4e being legitimate.

\subsection{Why this is called \emph{backward} induction}

The only equation in the recursion that contains no unknowns is Equation~\eqref{eq:bellman_terminal} --- the terminal one.  There is no future to integrate over; only the present belief and the present reward.  Once $V_K$ is a known function, Equation~\eqref{eq:bellman_general} at $k = K-1$ becomes solvable, because its right-hand side only references $V_K$.  That gives $V_{K-1}$, which feeds $V_{K-2}$, and so on, back to $V_0$.  The value function is constructed layer by layer, from the end of the experiment to the beginning.  This is ``backward induction on the horizon.''

\subsection{Value iteration is a different thing}

For completeness: Bellman's recursion as stated here is the \emph{finite-horizon} DP method.  The name ``value iteration'' usually refers to its infinite-horizon cousin, in which one repeatedly applies a Bellman operator to a single function $V$ on beliefs and waits for convergence to a fixed point.  Convergence there is by contraction (discount factor $< 1$).  In the finite-horizon case there is no iteration to convergence --- the terminal condition anchors the recursion and $K$ backward sweeps suffice exactly.  All three case studies in the companion paper are finite-horizon, and so is this tutorial.

\subsection{Where the computation goes}

In principle, the recursion is a finite calculation: there are $K$ layers and each layer is a max over actions of an expectation over observations.  In practice the obstruction is that $\bb$ is a continuous object (a probability density).  You cannot tabulate $V_k$ on the infinite space of possible beliefs.  Some reduction is needed:

\begin{itemize}[nosep, leftmargin=*]
\item A sufficient statistic of $\bb$ that lives in a finite or bounded space.  (The qubit example does this.)
\item A finitely parametrized model family for $\bb$, e.g.\ Gaussian moments.  (Case~C of the paper does this; the tutorial does not pursue it.)
\item A discrete state space in which $\bb$ is already a finite vector.  (The radar example.)
\end{itemize}

The two worked examples that follow show the first and third of these in detail.

\section{Example 1: Qubit flux sensor (continuous state, count-tuple reduction)}
\label{sec:qubit}

\subsection{Problem statement}

An external magnetic flux $\Phi_\mathrm{ext}$ threads a SQUID loop.  Its unknown value $\bx = \Phi_\mathrm{ext}$ is drawn from a uniform prior on $[0, \phi_{\max}\, \Phi_0]$ with $\phi_{\max} = 0.1$.  At each of $K = 4$ epochs we pick a Ramsey delay $\tau_j$ from a fixed menu of ten log-spaced delays in $[10, 320]$~ns and a repetition count $n \in \{1, 10\}$, then take $n$ Ramsey shots at that delay.  The likelihood for a single shot is
\begin{equation}
p(\by = 1 \mid \bx, \tau;\, \bc) \;=\; \tfrac{1}{2} + \tfrac{1}{2}\,\mathrm{th}(\bx, \bc)\, e^{-A(\bx, \bc)\,\tau - B(\bx, \bc)^2\, \tau^2}\, \cos\bigl[\Delta\omega(\bx, \bc)\, \tau\bigr],
\label{eq:ramsey}
\end{equation}
where $\mathrm{th}, A, B, \Delta\omega$ are closed-form expressions in $\bx$ and the hardware $\bc$.  The outcome of $n$ shots at a fixed delay is Binomial$(n, p)$.  The reward is negative terminal posterior variance, $\Phi_0(\bb_K) = -\mathrm{Var}(\bx \mid \bb_K)$.

\subsection{The measurement history reduces to a count-tuple}

The raw measurement history after $k$ epochs is the list
\begin{equation}
\mathcal{H}_k \;=\; \bigl( (\tau^{(1)}, y^{(1)}),\ (\tau^{(2)}, y^{(2)}),\ \ldots \bigr),
\end{equation}
one entry per individual shot, recording which delay it was taken at and whether it returned a $1$.  $\mathcal{H}_k$ can be very long (up to forty entries for $K = 4$ at $n = 10$ per epoch) and it contains redundancy.  The redundancy comes from two structural facts about the Ramsey likelihood, Equation~\eqref{eq:ramsey}:

\begin{enumerate}[nosep, leftmargin=*]
\item \emph{Conditional independence.}  Given $\bx$ and the hardware $\bc$, different shots are independent.  The likelihood of the entire history factors into a product over shots.
\item \emph{Conditional identity within a delay.}  Given $\bx$, all shots at the same delay $\tau_j$ have the same success probability $p(\bx, \tau_j;\, \bc)$.  They are therefore not merely independent but i.i.d.\ within each delay bin.
\end{enumerate}

Together these facts let us evaluate the likelihood of $\mathcal{H}_k$ by counting.  Let $N_j$ be the cumulative number of shots ever taken at delay $\tau_j$ in $\mathcal{H}_k$, and $M_j$ the cumulative number of those shots that returned a $1$.  Then
\begin{equation}
p(\mathcal{H}_k \mid \bx;\, \bc) \;=\; \prod_{j=1}^J p(\bx, \tau_j;\, \bc)^{M_j}\, \bigl[1 - p(\bx, \tau_j;\, \bc)\bigr]^{N_j - M_j}.
\label{eq:likelihood_counts}
\end{equation}
Every other aspect of the history --- the order of the shots, how they were split across epochs, how $n$ was partitioned between $1$-shot and $10$-shot bursts --- drops out of Equation~\eqref{eq:likelihood_counts}.

The pair $(\mathbf{N}, \mathbf{M}) = (N_1, \ldots, N_J,\ M_1, \ldots, M_J)$ is, by the Fisher--Neyman factorization theorem, a \emph{sufficient statistic} of the history for inference about $\bx$: any two histories that agree on $(\mathbf{N}, \mathbf{M})$ are indistinguishable from the point of view of $\bx$.  We arrange them into a single $2J$-integer object called the \emph{count-tuple},
\begin{equation}
\mathbf{n}_k \;=\; \bigl(N_1, M_1,\ N_2, M_2,\ \ldots,\ N_J, M_J\bigr).
\label{eq:count_tuple}
\end{equation}
$\mathbf{n}_k$ is lossless (nothing about $\bx$ has been discarded), permutation-invariant (shot order does not appear), and bounded (each $N_j \le Kn_{\max} = 40$).

\subsection{The belief reduces automatically}

The belief at step $k$ is a functional of the history $\mathcal{H}_k$ through Bayes' rule:
\begin{equation}
\bb_k(\bx) \;=\; \frac{p_0(\bx)\, p(\mathcal{H}_k \mid \bx;\, \bc)}{\int p_0(\bx')\, p(\mathcal{H}_k \mid \bx';\, \bc)\, d\bx'}.
\end{equation}
But the numerator depends on $\mathcal{H}_k$ only through Equation~\eqref{eq:likelihood_counts}, which depends only on $\mathbf{n}_k$.  Consequently
\begin{equation}
\bb_k(\bx) \;=\; \bb(\bx \mid \mathbf{n}_k;\, \bc) \;\propto\; p_0(\bx) \prod_{j=1}^J p(\bx, \tau_j;\, \bc)^{M_j}\,\bigl[1 - p(\bx, \tau_j;\, \bc)\bigr]^{N_j - M_j}.
\label{eq:belief_from_counts}
\end{equation}
The belief is a deterministic function of the count-tuple: once you know $\mathbf{n}$, you know $\bb$.  Two histories with the same count-tuple have the same belief, regardless of how the measurements were ordered or split.

\subsection{The DP table reduces for free}

Now the Bellman machinery inherits the reduction directly.  $V_k$ is a function of the belief; the belief is a function of the count-tuple; therefore $V_k$ is a function of the count-tuple: $V_k(\bb) = V_k(\mathbf{n})$.  The tabulation space collapses from the infinite-dimensional space of densities on $[0, \phi_{\max} \Phi_0]$ to the space of $2J$-integer tuples with bounded entries.

The chain of reductions is:
\begin{equation*}
\underbrace{\mathcal{H}_k}_{\text{raw history, long list}}\ \longrightarrow\ \underbrace{\mathbf{n}_k}_{\text{count-tuple, } 2J \text{ integers}}\ \longrightarrow\ \underbrace{\bb_k}_{\text{posterior, density}}\ \longrightarrow\ \underbrace{V_k(\mathbf{n}_k)}_{\text{value function, one scalar per tuple}}.
\end{equation*}
The first arrow is the Fisher--Neyman reduction on the likelihood; the second arrow is Bayes' rule; the third arrow is the Bellman recursion.  Critically, the first arrow is the \emph{only} reduction; the rest are functional consequences.  This is the right way to think of it --- not that the belief miraculously compresses, but that the \emph{data} compresses because of a symmetry of the likelihood, and the belief (being a function of the data) inherits the compression automatically.

\subsection{Reachability and the size of the DP table}

The count-tuple $\mathbf{n}_k$ is bounded: at $K = 4$ epochs with at most $n = 10$ shots each, the total shot count $\sum_j N_j$ is at most $40$ and the tuple is sparse (at most four bins touched, because there are at most four epochs).  The set of reachable count-tuples across all four epochs, as enumerated by a breadth-first search from the empty tuple, is approximately $10^7$ distinct tuples.  Each entry is a 64-bit float (the value-to-go).  The table fits in memory.

\begin{figure}[h]
\centering
\includegraphics[width=\textwidth]{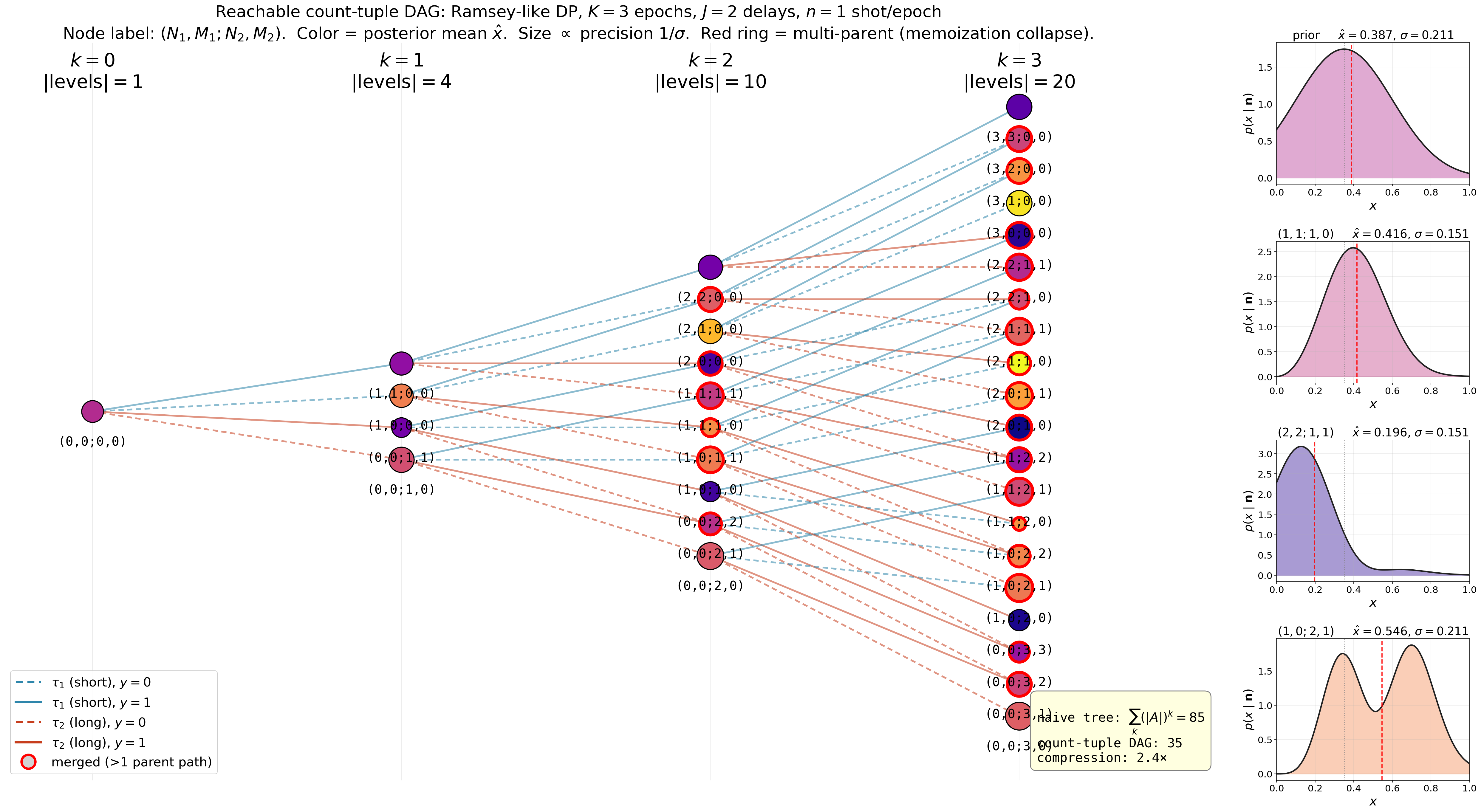}
\caption{Reachable count-tuple DAG for a Ramsey-like DP at $K = 3$ epochs, $J = 2$ delays, $n = 1$ shot per epoch.  \emph{Nodes:} reachable count-tuples $(N_1, M_1; N_2, M_2)$, one column per epoch; colour = posterior mean $\hat x$ (plasma scale); size $\propto$ posterior precision $1 / \sigma$; red ring marks nodes reached by more than one parent path (the memoization collapse).  \emph{Edges:} colour by action ($\tau_1$ / $\tau_2$), style by outcome ($y = 0$ dashed / $y = 1$ solid).  The compression-ratio box in the lower right compares the naive playout-tree size to the count-tuple DAG size.  \emph{Right column:} four representative posteriors --- the prior at the root and three distinct terminal-belief shapes at $k = 3$.}
\label{fig:belief_tree}
\end{figure}

Figure~\ref{fig:belief_tree} visualizes this enumeration at a very small problem size ($K = 3$, $J = 2$, $n = 1$): $1 \to 4 \to 10 \to 20$ reachable count-tuples across the four layers, $35$ nodes total, against a naive playout tree of $85$ leaves-plus-internal-nodes.  Even at this toy scale, $22$ of the $35$ count-tuples arise from more than one parent path (marked with a red ring) --- the memoization collapse that the count-tuple representation exploits.  At the paper's full scale ($K = 4$, $J = 10$, $L = 2$) the compression ratio grows to roughly two orders of magnitude.

\subsection{Backward induction on count-tuples, in detail}

\paragraph{Posterior from a count-tuple.}  To evaluate any expectation the DP needs, we must be able to turn a count-tuple into a posterior.  Discretize $\bx$ on a grid of $K_\Phi = 256$ points $\bx^{(1)}, \ldots, \bx^{(K_\Phi)}$ uniformly spaced across $[0, \phi_{\max} \Phi_0]$.  The posterior at the $i$-th grid point, given count-tuple $\mathbf{n}$, is
\begin{equation}
\bb(\bx^{(i)} \mid \mathbf{n}) \;\propto\; p_0(\bx^{(i)}) \prod_{j=1}^J p(\bx^{(i)}, \tau_j; \bc)^{M_j}\, \bigl[1 - p(\bx^{(i)}, \tau_j; \bc)\bigr]^{N_j - M_j},
\label{eq:posterior_from_counts}
\end{equation}
with the constant of proportionality fixed by $\sum_i \bb(\bx^{(i)} \mid \mathbf{n}) = 1$.  The posterior variance and the outcome probabilities needed below are both computed from Equation~\eqref{eq:posterior_from_counts} on the fly.

\paragraph{Marginal outcome probability.}  If the current posterior (given $\mathbf{n}$) is $\bb(\bx \mid \mathbf{n})$ and we take an action $(\tau, n)$, the distribution over the number of ``1''s is
\begin{equation}
\Pr(m \mid \mathbf{n}, \tau, n) \;=\; \sum_i \bb(\bx^{(i)} \mid \mathbf{n})\; \mathrm{Binomial}\bigl(m \,\big|\, n,\ p(\bx^{(i)}, \tau; \bc)\bigr).
\label{eq:marginal_outcome}
\end{equation}
This is the marginal likelihood that will serve as the weighting in the Bellman backup's expectation.

\paragraph{Child tuple.}  If we take $(\tau_j, n)$ and observe $m$ ``1'' outcomes, the new count-tuple has the same entries as $\mathbf{n}$ except $(N_j, M_j)$ becomes $(N_j + n,\, M_j + m)$.  We write this $\mathbf{n}'(\mathbf{n}, j, n, m)$.

\paragraph{Bellman backup, derived step by step.}  Equations~\eqref{eq:qubit_terminal} and~\eqref{eq:qubit_backup} below are the general Bellman recursion of \S\ref{sec:bellman_principle} (Equations~\eqref{eq:bellman_terminal}--\eqref{eq:bellman_general}) specialized to four qubit-specific facts: the reward, the action space, the observation reduction, and the sufficient statistic.  We articulate each substitution explicitly so the derivation can be verified line by line.

\textit{Step 1: choose the reward $\Phi$.}  Section~\ref{sec:blocks} introduced three families of pointwise rewards.  For the qubit problem we use the Bayesian-MSE-aligned reward, with terminal expression $\Phi_0(\bx, \bb_K) = -\mathrm{Var}(\bx \mid \bb_K)$.  This choice is motivated by the tower-law identity $\E[(\hat\bx_{\text{pm}} - \bx)^2] = \E[\mathrm{Var}(\bx \mid \bb_K)]$: maximizing the prior-averaged $\Phi_0$ is identically minimizing the deployed Bayesian MSE of the posterior-mean estimator.

\textit{Step 2: collapse the leaf expectation.}  Substituting Step 1 into the general terminal Equation~\eqref{eq:bellman_terminal}, the integrand $-\mathrm{Var}(\bx \mid \bb_K)$ is a deterministic function of the belief $\bb_K$ (it does not depend on the realization of $\bx$ at the leaf).  The conditional expectation is therefore trivial:
\begin{equation*}
V_K(\bb) \;=\; \E_{\bx \mid \bb}\!\bigl[-\mathrm{Var}(\bx \mid \bb)\bigr] \;=\; -\mathrm{Var}(\bx \mid \bb).
\end{equation*}
This collapse is special to $\Phi_{\text{Var}}$; it would not happen for $\Phi_{\text{MI}}$, whose leaf integrand retains an explicit $\bx$-dependence inside the entropy expectation.

\textit{Step 3: encode the belief by the count-tuple.}  Equation~\eqref{eq:belief_from_counts} established that the posterior $\bb_k(\bx)$ depends on the history only through the count-tuple $\mathbf{n} = (N_j, M_j)_{j=1}^J$.  Writing the terminal belief as $\bb(\cdot \mid \mathbf{n})$ converts $V_K(\bb)$ into a function of the count-tuple,
\begin{equation*}
V_K(\mathbf{n}) \;=\; -\mathrm{Var}\bigl(\bx \mid \mathbf{n}\bigr),
\end{equation*}
which is Equation~\eqref{eq:qubit_terminal} below.

\textit{Step 4: specialize the action.}  In the general recursion the outer $\max_{\bs \in \mathcal{S}}$ ranges over an abstract action space.  In the qubit sensor every action is the pair $\bs = (\tau_j, n)$ with delay-index $j \in \{1, \ldots, 10\}$ and shot count $n \in \{1, 10\}$.  The outer max becomes a finite enumeration of $J \times L = 20$ candidates,
\begin{equation*}
\max_{\bs \in \mathcal{S}} \;\longrightarrow\; \max_{\substack{j \in \{1, \ldots, 10\}\\ n \in \{1, 10\}}}.
\end{equation*}

\textit{Step 5: collapse the observation expectation $\by \to m$.}  At fixed action $(\tau_j, n)$, the raw observation $\by_k$ is the bit string of $n$ Bernoulli outcomes.  The Ramsey likelihood (Equation~\eqref{eq:likelihood_counts}) depends on the bit string only through the count $m \in \{0, 1, \ldots, n\}$ of ``1'' outcomes, by the same i.i.d.-within-a-delay argument that gave the count-tuple sufficient statistic.  The expectation over the $2^n$-outcome bit string therefore reduces to an $(n+1)$-element finite sum,
\begin{equation*}
\E_{\by_k \mid \bb, (\tau_j, n)}[f(\by_k)] \;=\; \sum_{m=0}^{n}\, \Pr(m \mid \mathbf{n}, \tau_j, n)\; f(m),
\end{equation*}
with $\Pr(m \mid \mathbf{n}, \tau_j, n)$ given by Equation~\eqref{eq:marginal_outcome}.  This reduction is loss-less: no information about the bit-string-internal ordering enters the recursion because none enters the likelihood.

\textit{Step 6: replace the belief operator $T$ by the count-tuple update.}  The general operator $T(\bb, \bs, \by;\, \bc)$ is Bayes' rule applied to belief $\bb$ given action $\bs$ and observation $\by$.  Applied to a count-tuple-encoded belief at action $(\tau_j, n)$ and observed count $m$, Bayes' rule produces a new belief encoded by the count-tuple
\begin{equation*}
\mathbf{n}'(\mathbf{n}, j, n, m) \;=\; \bigl(\ldots,\, (N_{j-1}, M_{j-1}),\, (N_j + n,\, M_j + m),\, (N_{j+1}, M_{j+1}),\, \ldots\bigr),
\end{equation*}
with only the $j$-th entry incremented and all other entries untouched.  The recursive lookup $V_{k+1}(T(\bb, \bs, \by;\, \bc))$ becomes the table lookup $V_{k+1}(\mathbf{n}'(\mathbf{n}, j, n, m))$ on a finite, memoized table.

\textit{Step 7: assemble.}  Substituting Steps 4, 5, and 6 into the general backup Equation~\eqref{eq:bellman_general} and Step 3 into the general terminal Equation~\eqref{eq:bellman_terminal} yields the qubit-specific recursion
\begin{align}
V_K(\mathbf{n}) &\;=\; -\mathrm{Var}\bigl(\bx \mid \mathbf{n}\bigr), \label{eq:qubit_terminal}\\
V_k(\mathbf{n}) &\;=\; \max_{\substack{j \in \{1, \ldots, 10\}\\ n \in \{1, 10\}}}\;
\sum_{m = 0}^{n}
\Pr(m \mid \mathbf{n}, \tau_j, n)\; V_{k+1}\bigl(\mathbf{n}'(\mathbf{n}, j, n, m)\bigr),
\label{eq:qubit_backup}
\end{align}
for $k = K - 1, \ldots, 0$.  The optimal adaptive policy is $\pi^\star_k(\mathbf{n}) = (\tau_{j^\star}, n^\star)$, the argmax of Equation~\eqref{eq:qubit_backup}.  The joint-DP value at the prior is $V^\star = V_0(\mathbf{0})$, evaluated at the all-zero count-tuple representing the uniform prior.

\paragraph{What each step relied on.}  Step 1 relied on the modeling choice $\Phi = \Phi_{\text{Var}}$.  Step 2 relied on the leaf integrand being a function of $\bb_K$ alone, special to $\Phi_{\text{Var}}$ (no analogous collapse exists for $\Phi_{\text{MI}}$ or $\Phi_{\log J}$).  Steps 3 and 6 relied on the Fisher--Neyman count-tuple reduction of Equation~\eqref{eq:likelihood_counts}.  Steps 4 and 5 relied on the discrete action menu and the binomial-counted observation specific to the Ramsey protocol.  Each substitution is exact; no approximation has entered.  The resulting recursion is therefore the exact Bellman recursion for the qubit sensor, expressed in terms that admit memoized tabulation on count-tuple keys.

\subsection{A miniature trace}

To make Equations~\eqref{eq:qubit_terminal}--\eqref{eq:qubit_backup} concrete, shrink the problem: $K = 2$, $J = 2$ delays, $n \in \{1\}$ only.  There are four candidate actions total, $(\tau_1, 1), (\tau_2, 1)$ repeated at each epoch; twelve distinct child tuples at epoch $2$ (two delays times six possible aggregate counts of zero or one across the two shots --- but since $n=1$ at each step the reachable tuples after one epoch are $\{(1,0,0,0), (1,1,0,0), (0,0,1,0), (0,0,1,1)\}$, four of them, each representing a different shot location and outcome).  Starting from the all-zero tuple:

\begin{enumerate}[nosep, leftmargin=*]
\item \emph{Terminal layer $k = 2$.}  For each of the at most $4 + 4 \times 4 = 20$ reachable tuples, compute the posterior variance from Equation~\eqref{eq:posterior_from_counts} and store $V_2(\mathbf{n}) = -\mathrm{Var}$.
\item \emph{Layer $k = 1$.}  For each of the four tuples reachable after one epoch, try both remaining actions.  For each, compute the two child tuples (outcomes $m = 0$ and $m = 1$), look up their $V_2$ values, and compute the marginal-outcome-weighted average.  Take the best of the two actions.
\item \emph{Root $k = 0$.}  Start at $\mathbf{n} = \mathbf{0}$.  Try both actions.  For each, compute the two child tuples at $k = 1$, look up their $V_1$, and average.  Take the best.  The result is $V^\star$ at the initial belief; the argmax tells us the optimal first action.
\end{enumerate}

For $K = 4$, $J = 10$, $n \in \{1, 10\}$ the identical loop runs over $\sim 10^7$ tuples and takes minutes on a single thread (the DP is embarrassingly memoized, each tuple visited once).  The logic is unchanged.

\subsection{What the count-tuple reduction bought us}

Without the reduction, a brute-force playout tree at $K = 4$, $|\mathcal{S}| = 20$, and up to $11$ outcomes per action would have $(20 \times 11)^4 \approx 2 \times 10^9$ leaves.  With the reduction, a permutation of playouts that produces the same count-tuple collapses to one table entry; the reachable state space shrinks by roughly two orders of magnitude.  The exactness of the reduction --- no information about $\bx$ is lost by representing a history as counts --- is what makes this a \emph{lossless} speedup rather than an approximation.

\section{Example 2: Radar beam search (discrete state, finite-belief DP)}
\label{sec:radar}

\subsection{Problem statement}

A target sits in one of sixteen cells arranged on a ring: $\bx \in \{1, \ldots, 16\}$, with uniform prior $p_0(\bx) = 1/16$.  At each of $K = 4$ epochs we point a beam at some cell $\bs_k \in \{1, \ldots, 16\}$ and observe $\by_k \in \{0, 1\}$ with Bernoulli likelihood
\begin{equation}
p(\by_k = 1 \mid \bx, \bs_k;\, \bc) \;=\; p_\mathrm{FA} + (p_\mathrm{D} - p_\mathrm{FA}) \cdot G_{\bc}(\theta_{\bs_k} - \varphi_{\bx}),
\label{eq:radar_likelihood}
\end{equation}
with $p_\mathrm{D} = 0.9$, $p_\mathrm{FA} = 0.05$, cell bearings $\varphi_{\bx} = 2\pi(\bx - 1)/16$, and $G_{\bc}$ a top-hat beam pattern of integer width $m$ (the hardware design variable).  The reward is the mutual information $I(\bx; \by_{1:K})$, which equals the prior entropy minus the expected terminal posterior entropy; maximizing it is equivalent to minimizing the expected terminal posterior entropy.

\subsection{The belief is a probability vector}

Since $\bx$ is discrete with $|\mathcal{X}| = 16$, the belief $\bb_k$ is a probability vector in the $15$-simplex.  It is already finite-dimensional; no reduction is needed.  The Bayes update $T$ is elementary: at belief $\bb$, action $\bs$, outcome $\by$,
\begin{equation}
\bb'(\bx) \;=\; \frac{p(\by \mid \bx, \bs;\, \bc)\, \bb(\bx)}{\sum_{\bx'} p(\by \mid \bx', \bs;\, \bc)\, \bb(\bx')}.
\label{eq:radar_bayes}
\end{equation}

The DP is now a straight implementation of Equations~\eqref{eq:bellman_terminal}--\eqref{eq:bellman_general} on beliefs.

\subsection{Reachable beliefs and the DP table}

Here the tabulation strategy is not to enumerate all probability vectors (uncountably many) but to enumerate the \emph{reachable} beliefs --- those that can actually arise from the uniform prior by a sequence of allowed actions and allowed outcomes.  At $K = 4$ with binary outcomes and $|\mathcal{S}| = 16$, the number of reachable beliefs across all four epochs is bounded by $16^4 \times 2^4 = 10^6$; in practice many fewer because many action-outcome sequences lead to the same belief.  A breadth-first traversal from $\bb_0 = p_0$ records every reachable belief once and assigns it an index in a table.

Alternatively, in the special case of the top-hat beam, the belief at any point has at most a constant number of distinct values --- the cells inside the beam, the cells outside, conditioned on each detection history --- so the reachable belief has a closed-form structure.  But conceptually, the DP is just ``belief in, value out.''

\subsection{Backward induction on beliefs, in detail}

Let $\mathcal{B}_k$ be the set of beliefs reachable exactly at step $k$ (so $\mathcal{B}_0 = \{p_0\}$).

\paragraph{Terminal layer.}  For every $\bb \in \mathcal{B}_K$,
\begin{equation}
V_K(\bb) \;=\; \sum_{\bx = 1}^{16} \bb(\bx)\, \log \bb(\bx) - \text{(prior entropy constant)},
\label{eq:radar_terminal}
\end{equation}
the negative posterior entropy (plus a constant that cancels in the argmax).  This is the reward for that terminal belief.

\paragraph{Inductive step.}  For $k = K-1, K-2, \ldots, 0$ and each $\bb \in \mathcal{B}_k$:
\begin{equation}
V_k(\bb) \;=\; \max_{\bs \in \{1, \ldots, 16\}}\; \sum_{\by \in \{0, 1\}} \Pr(\by \mid \bb, \bs;\, \bc)\; V_{k+1}\bigl(T(\bb, \bs, \by;\, \bc)\bigr),
\label{eq:radar_backup}
\end{equation}
where $\Pr(\by \mid \bb, \bs;\, \bc) = \sum_{\bx} \bb(\bx)\, p(\by \mid \bx, \bs; \bc)$ is the marginal outcome probability under the current belief.  The argmax is the optimal adaptive action at belief $\bb$ with $K - k$ steps left.

\paragraph{Root.}  $V^\star = V_0(p_0)$.  The globally optimal policy is the collection $\{\pi^\star_k : \bb \mapsto \argmax_{\bs} \ldots\}$ for $k = 0, \ldots, K-1$.

\subsection{A miniature trace: one cell, one beam, \texorpdfstring{$K = 2$}{K=2}}

To see the structure of an optimal adaptive radar policy, consider the simplification $K = 2$ with a narrow top-hat beam that covers exactly one cell ($m = 1$).  From the uniform prior:

\begin{enumerate}[nosep, leftmargin=*]
\item At epoch $1$, for any choice of pointing cell $\bs_1$, all $16$ cells are equally plausible.  The Bellman backup says try every $\bs_1$; by symmetry they are all equivalent, so pick $\bs_1 = 1$ without loss of generality.
\item Two outcomes are possible.  If $\by_1 = 1$, the posterior concentrates on cell $1$ up to the false-alarm rate; the remaining cells are weighted by $p_\mathrm{FA} / [p_\mathrm{D} (1 - p_\mathrm{FA}) + \ldots]$.  If $\by_1 = 0$, the posterior shifts weight slightly away from cell $1$.
\item At epoch $2$, the Bellman backup tries every $\bs_2$.  After a detection at cell $1$, the best $\bs_2$ is to point away from cell $1$ at whichever cell now carries the second-highest belief (the MI gain is maximized by sampling from a cell you are uncertain about).  After a miss at cell $1$, the best $\bs_2$ is similarly to point elsewhere.
\end{enumerate}

In words: the argmax at epoch $2$ is \emph{different} depending on whether $\by_1 = 0$ or $\by_1 = 1$.  That is the adaptive gain.  A fixed schedule of $(\bs_1 = 1, \bs_2 = 2)$ commits to cell $2$ before seeing $\by_1$ and thus cannot exploit the information in the outcome.

For the full $K = 4$, $m = 9$ problem (the in-family optimum of the companion paper's Section~5), the Bellman backup produces an adaptive policy that behaves like binary search on the posterior support: the first beam splits the ring in half; each subsequent beam targets the half whose posterior weight is highest, narrowing by roughly a factor of two per step.

\section{What the two examples share, and where they differ}
\label{sec:comparison}

\begin{center}
\begin{tabular}{@{}lll@{}}
\toprule
 & \textbf{Qubit (Case~B)} & \textbf{Radar (Case~A)} \\
\midrule
State space $\mathcal{X}$           & continuous, $[0, \phi_{\max}\Phi_0]$  & discrete, $\{1, \ldots, 16\}$ \\
Belief object                       & density on $\mathcal{X}$              & probability vector in $\R^{16}$ \\
Belief reduction                    & count-tuple (20 integers)             & (none needed) \\
Action set $\mathcal{S}$            & 10 delays $\times$ 2 rep counts       & 16 pointing cells \\
Observation $\by$                   & Binomial$(n, p)$                       & Bernoulli \\
Outcome count per action            & 2 or 11                                & 2 \\
Horizon $K$                         & 4                                      & 4 \\
Reward                              & $-\mathrm{Var}(\bx \mid \bb_K)$       & $-H(\bb_K)$ (equiv.\ MI) \\
DP table size                       & $\sim 10^7$ count-tuples              & $\sim 10^6$ reachable beliefs \\
\bottomrule
\end{tabular}
\end{center}

The commonalities are load-bearing.  Both problems fit the POMDP template, both reward families are scalar functionals of the terminal belief, both Bellman recursions take the form of Equation~\eqref{eq:bellman_general}, and both are solved exactly by backward induction.  The divergences are in the shape of the belief and the size of the DP table, both of which are consequences of the state-space dimensionality.

Two lessons worth taking away:

\begin{enumerate}[nosep, leftmargin=*]
\item \textbf{Reduction is essential in continuous-state problems.}  The qubit problem only becomes computable because the Ramsey likelihood has a permutation symmetry that collapses the posterior to a 20-integer tuple.  A generic continuous-state POMDP has no such luck and must be relaxed --- Gaussian-moment propagation (EKF), particle filters, or neural-belief-state methods --- at which point the DP is no longer exact.
\item \textbf{Exactness is cheap in discrete-state problems.}  The radar problem runs directly on the natural probability-vector belief and returns a provably optimal policy without approximations, as long as the reachable belief set fits in memory.  At $K = 4$, $|\mathcal{X}| = 16$ it does; at $K = 20$ or $|\mathcal{X}| = 256$ it would not, and the same relaxation pathway as the qubit problem would be needed.
\end{enumerate}

\section{Belief augmentation: when the posterior is not enough}
\label{sec:augmentation}

Both examples in this tutorial share a quiet simplification: the reward is a functional of the terminal posterior $\bb_K$ alone.  In the qubit problem $-\mathrm{Var}(\bx \mid \bb_K)$ depends only on the shape of the posterior, and in the radar problem so does $-H(\bb_K)$.  Once the posterior is determined, the reward is determined --- no reference to which actions produced it.  This is what lets Step~2 of §\ref{sec:derivation} go through: $\bb_K$ serves as a sufficient statistic of the trajectory \emph{for the reward}, not just for $\bx$.

Not every sensing reward has this property.  The main paper's Case~C uses the Fisher-information-aligned reward
\begin{equation}
\Phi(\bx, \bs_{1:K};\, \bc) \;=\; \tfrac{1}{2}\, \log \det J_N(\bx, \bs_{1:K};\, \bc),
\label{eq:phi_logJ_tutorial}
\end{equation}
where the accumulated Fisher-information matrix is
\begin{equation}
J_N(\bx, \bs_{1:K};\, \bc) \;=\; J_0 + \sum_{k=1}^K n_k\, J_F(\bx, \tau_k;\, \bc),
\end{equation}
and $J_F$ is the per-shot Fisher matrix.  The issue is immediate: $J_N$ depends \emph{explicitly} on the action trajectory $\bs_{1:K} = ((\tau_1, n_1), \ldots)$, not just on the posterior those actions produced.  Two schedules that happen to land on the same $\bb_K$ via different $\{(\tau_k, n_k)\}$ choices give different $J_N$, hence different reward.

In other words: for $\Phi_{\log J}$, $\bb_K$ is \emph{not} a sufficient statistic of the trajectory for the reward.  Step~2 of the derivation in §\ref{sec:derivation} fails on its face.  A direct application of Bellman's recursion on $\bb_k$ alone would miscompute the value function --- it would forget the schedule-dependent accumulator.

\subsection{The fix: enlarge what ``belief'' means}

The cure is to redefine the state of the DP so that whatever extra information the reward requires rides along.  For $\Phi_{\log J}$ we define the \emph{augmented belief}
\begin{equation}
\tilde{\bb}_k \;\coloneqq\; \bigl(\bb_k,\ J_k\bigr), \qquad J_k \;\coloneqq\; J_0 + \sum_{j=1}^k n_j\, J_F(\bx, \tau_j;\, \bc).
\label{eq:augmented_belief}
\end{equation}
Two quantities now move together: the posterior over $\bx$, which still updates by Bayes' rule, and the accumulated Fisher matrix, which updates by additive accumulation:
\begin{equation}
J_{k+1} \;=\; J_k + n_{k+1}\, J_F(\bx, \tau_{k+1};\, \bc).
\end{equation}
The augmented update operator is
\begin{equation}
\tilde{T}\bigl(\tilde{\bb}, \bs, \by;\, \bc\bigr) \;=\; \bigl(\, T(\bb, \bs, \by;\, \bc),\ J + n\, J_F(\bx, \tau;\, \bc) \,\bigr),
\end{equation}
with $\bs = (\tau, n)$.  The reward at the terminal augmented belief is a pure functional of $\tilde{\bb}_K$:
\begin{equation}
\tilde{\Phi}_0(\tilde{\bb}_K;\, \bc) \;=\; \E_{\bx \mid \bb_K}\!\bigl[\tfrac{1}{2}\, \log \det J_K\bigr].
\end{equation}
With $(\bb, T, \Phi_0)$ replaced by $(\tilde{\bb}, \tilde{T}, \tilde{\Phi}_0)$, the four-step derivation of §\ref{sec:derivation} goes through unchanged --- in particular, $\tilde{\bb}_K$ \emph{is} a sufficient statistic of the trajectory for the augmented reward, so Step~2 is rescued.  Bellman's recursion now lives on $\tilde{\bb}$, and backward induction produces the correct $V_k(\tilde{\bb};\, \bc)$.

\subsection{What augmentation costs and what it buys}

The cost is bookkeeping: for a $d$-dimensional state $\bx$, $J_k$ is a symmetric $d \times d$ matrix, carrying $d(d+1)/2$ extra real numbers in the state.  In exact DP this multiplies the effective state-space dimensionality and the table size.  In moment-based DP (Case~C of the paper), where $\bb_k$ is already a compact Gaussian summary, the augmentation by $J_k$ is a mild expansion of a small tuple.

The benefit is that the Bellman machinery proceeds identically to the unaugmented case.  The value function $V_k(\tilde{\bb};\, \bc)$ is well-posed, the recursion is exact, the argmax produces an optimal adaptive policy, and (importantly) the envelope-theorem outer gradient on $\bc$ that the companion paper exploits applies verbatim to the augmented belief.

\subsection{A rule of thumb for when to augment}

Look at the pointwise reward.  List every action-dependent quantity it references.  For each, ask: is it a functional of the posterior $\bb_K$ alone?  If yes, leave the belief unaugmented; if no, augment by whichever accumulator would make the reward a functional of the augmented belief.  Common triggers:

\begin{itemize}[nosep, leftmargin=*]
\item \textbf{Fisher-information rewards.}  $\log \det J_N$, $\mathrm{tr}\, J_N$, any function of an action-accumulated Fisher matrix.  Augment by $J_k$.
\item \textbf{Cumulative cost or budget constraints.}  A reward that penalizes $\sum_k c(\bs_k)$ --- shot-count budgets, power budgets, time-resource constraints.  Augment by the running cost sum.
\item \textbf{Path-dependent rewards.}  Rewards that integrate observations or actions over the trajectory (entropy production, work extraction, dwell-time penalties).  Augment by the relevant path integral.
\item \textbf{Constraint indicators.}  ``Has the trajectory stayed in a safe set?''  Augment by a binary safety flag that latches off the first time the trajectory leaves the set.
\end{itemize}

The recipe is always the same: redefine the DP's state as (posterior, running accumulator), so that the reward is a functional of the new state and the Bayesian update $T$ is closed over the new state.  The tutorial's qubit and radar examples correspond to the trivial case where no accumulator is needed.

\section{Connection to joint hardware-policy co-design}
\label{sec:joint_dp}

Everything in this tutorial so far has assumed the hardware $\bc$ is fixed: the DP produces the optimal adaptive policy $\pi^\star(\bc)$ for whichever hardware was handed in.  The companion paper wraps this inner DP inside an outer optimization over $\bc$ --- gradient descent on
\begin{equation}
V^\star(\bc) \;\coloneqq\; V_0(p_0;\, \bc).
\end{equation}
This section describes how the outer gradient $\partial V^\star / \partial \bc$ is computed, and why --- despite the Bellman recursion being a max-of-expectations nested $K$ levels deep --- the gradient comes out of ordinary reverse-mode automatic differentiation with a single structural trick.

\subsection{Three enabling observations}

\begin{itemize}[nosep, leftmargin=*]
\item \textbf{Envelope theorem.}  The derivative of $V^\star(\bc)$ at a point where the inner policy is Bellman-optimal does not require differentiating through the argmax.  The inner policy is treated as a constant function of $\bc$ when taking the outer gradient; only the \emph{explicit} $\bc$-dependence of the reward and the belief update matters.
\item \textbf{Policy-iteration warm-starting.}  As $\bc$ drifts under outer gradient steps, the previously computed $\pi^\star(\bc_{\mathrm{old}})$ is near-optimal for $\bc_{\mathrm{new}}$.  A full rebuild of the inner DP is not needed every outer step; a partial update suffices.
\item \textbf{Decoupling of problem sizes.}  The cost of the outer loop scales with $d_c$ (the dimension of $\bc$); the cost of the inner loop scales with the DP table size.  These are largely independent, which is why $d_c \sim 10^5$ (the photonic case) can coexist with inner DPs that fit in memory.
\end{itemize}

\subsection{Differentiating the Bellman recursion through \texorpdfstring{$\bc$}{c}}

The central obstruction to a mechanical gradient calculation is the $\max$ in the Bellman backup, Equation~\eqref{eq:bellman_general}.  The $\max$ is non-smooth, and a naive differentiation through it would pick up subgradient contributions corresponding to how the argmax itself shifts with $\bc$.  The envelope theorem says those contributions vanish at the optimum.  We spell out what remains.

\paragraph{The envelope theorem, formally.}  Define the ``Q-function'' at step $k$,
\begin{equation}
Q_k(\bb, \bs;\, \bc) \;\coloneqq\; \E_{\by_k \mid \bb, \bs, \bc}\!\bigl[V_{k+1}(T(\bb, \bs, \by_k;\, \bc);\, \bc)\bigr],
\label{eq:Q_def}
\end{equation}
so that $V_k(\bb;\, \bc) = \max_{\bs_k} Q_k(\bb, \bs_k;\, \bc)$ and $\bs_k^\star(\bb;\, \bc) = \argmax_{\bs_k} Q_k(\bb, \bs_k;\, \bc)$.  The total derivative of $V_k$ with respect to $\bc$ splits into an indirect piece (via $\bs_k^\star$) and a direct piece:
\begin{equation}
\frac{\partial V_k}{\partial \bc}(\bb;\, \bc)
\;=\; \underbrace{\frac{\partial Q_k}{\partial \bs_k}(\bb, \bs_k^\star;\, \bc) \cdot \frac{\partial \bs_k^\star}{\partial \bc}(\bb;\, \bc)}_{= 0 \ \text{at the argmax}}
\;+\; \frac{\partial Q_k}{\partial \bc}(\bb, \bs_k^\star;\, \bc).
\end{equation}
The first term vanishes because $\partial Q_k / \partial \bs_k = 0$ at an interior argmax (the first-order optimality condition); what survives is just the partial derivative of $Q_k$ with $\bs_k$ held fixed at $\bs_k^\star$:
\begin{equation}
\boxed{\ \frac{\partial V_k}{\partial \bc}(\bb;\, \bc) \;=\; \frac{\partial Q_k}{\partial \bc}(\bb, \bs_k^\star;\, \bc).\ }
\label{eq:envelope_tutorial}
\end{equation}
Equation~\eqref{eq:envelope_tutorial} is the operational form of the envelope theorem: to differentiate $V_k$ with respect to $\bc$, freeze the argmax and differentiate the integrand.  No argmax sensitivity is needed.

\paragraph{Three contributions to $\partial Q_k / \partial \bc$.}  Write $Q_k$ as an explicit sum over outcomes (discrete $\by_k$ for the qubit and radar problems, integral otherwise):
\begin{equation}
Q_k(\bb, \bs_k^\star;\, \bc) \;=\; \sum_{\by_k}\, \Pr(\by_k \mid \bb, \bs_k^\star, \bc)\, V_{k+1}\bigl(T(\bb, \bs_k^\star, \by_k;\, \bc);\, \bc\bigr).
\end{equation}
Applying the product and chain rules with $\bs_k^\star$ held fixed,
\begin{align}
\frac{\partial Q_k}{\partial \bc} \;=\; \sum_{\by_k} \biggl[ &\underbrace{\frac{\partial \Pr(\by_k \mid \bb, \bs_k^\star, \bc)}{\partial \bc}\, V_{k+1}\bigl(T(\cdots);\, \bc\bigr)}_{\text{(i) outcome-probability term}} \notag \\
&\; +\; \Pr(\by_k \mid \cdots)\, \underbrace{\frac{\partial V_{k+1}}{\partial \bb}\bigl(T;\, \bc\bigr) \cdot \frac{\partial T(\bb, \bs_k^\star, \by_k;\, \bc)}{\partial \bc}}_{\text{(ii) belief-update term}} \notag \\
&\; +\; \Pr(\by_k \mid \cdots)\, \underbrace{\frac{\partial V_{k+1}}{\partial \bc}\bigl(T;\, \bc\bigr)}_{\text{(iii) recursive term}} \biggr].
\label{eq:dV_unfolded}
\end{align}
Three pieces of the $\bc$-dependence enter:
\begin{enumerate}[nosep, leftmargin=*]
\item \textbf{Outcome-probability (i).}  $\bc$ changes the likelihood of the measurement $\by_k$, which re-weights the future continuations.
\item \textbf{Belief-update (ii).}  $\bc$ enters the Bayesian update $T$ through the likelihood factor, so the posterior landed on after observing $\by_k$ shifts with $\bc$.
\item \textbf{Recursive (iii).}  $V_{k+1}$ itself depends on $\bc$ --- this piece is picked up by the same formula applied at step $k+1$, which is the recursion that gives the gradient its backward-induction structure.
\end{enumerate}
The terminal gradient is $\partial V_K / \partial \bc = \partial \Phi_0^{\mathrm{belief}}(\bb;\, \bc) / \partial \bc$, obtained by differentiating the reward at the terminal belief with respect to its own $\bc$ arguments.

Equation~\eqref{eq:dV_unfolded} mirrors the Bellman recursion itself: same DP graph, same memoization, same backward order.  If $V$ fits in memory, so does $\partial V / \partial \bc$.

\subsection{A parallel recursion for \texorpdfstring{$\partial V_k / \partial \bb$}{dV\_k/db}}
\label{sec:dV_db}

Equation~\eqref{eq:dV_unfolded} contains a factor $\partial V_{k+1} / \partial \bb$ inside the belief-update term (ii).  Conceptually this is a gradient of the value function with respect to its belief argument, and it satisfies its own backward recursion --- a cleanly analogous story to the $\partial V_k / \partial \bc$ one.  We derive it here because the derivation is short and makes explicit what the belief-update term in Equation~\eqref{eq:dV_unfolded} is actually using.

\paragraph{Envelope identity for $\bb$.}  The same envelope-theorem argument applies with $\bb$ in place of $\bc$.  Differentiating $V_k(\bb; \bc) = Q_k(\bb, \bs_k^\star(\bb; \bc); \bc)$ with respect to $\bb$,
\begin{equation}
\frac{\partial V_k}{\partial \bb}(\bb;\, \bc) \;=\; \underbrace{\frac{\partial Q_k}{\partial \bs_k}(\bb, \bs_k^\star; \bc) \cdot \frac{\partial \bs_k^\star}{\partial \bb}(\bb; \bc)}_{= 0 \text{ at the argmax}} \;+\; \frac{\partial Q_k}{\partial \bb}(\bb, \bs_k^\star; \bc).
\end{equation}
The indirect piece vanishes by first-order optimality at $\bs_k^\star$, and the direct partial is what we need:
\begin{equation}
\frac{\partial V_k}{\partial \bb}(\bb;\, \bc) \;=\; \frac{\partial Q_k}{\partial \bb}(\bb, \bs_k^\star; \bc).
\label{eq:env_identity_b}
\end{equation}

\paragraph{Unfolding.}  Write $Q_k$ as a sum over outcomes (as in §\ref{sec:derivation}),
\begin{equation}
Q_k(\bb, \bs_k^\star;\, \bc) \;=\; \sum_{\by_k}\, \Pr(\by_k \mid \bb, \bs_k^\star, \bc)\, V_{k+1}\bigl(T(\bb, \bs_k^\star, \by_k;\, \bc);\, \bc\bigr),
\end{equation}
and differentiate with respect to $\bb$ using the product and chain rules:
\begin{align}
\frac{\partial V_k}{\partial \bb}(\bb;\, \bc) \;=\; \sum_{\by_k} \biggl[\,
&\underbrace{\frac{\partial \Pr(\by_k \mid \bb, \bs_k^\star, \bc)}{\partial \bb}\, V_{k+1}\bigl(T(\cdots);\, \bc\bigr)}_{\text{(I) outcome-probability through }\bb} \notag\\
& +\; \Pr(\by_k \mid \cdots)\, \underbrace{\frac{\partial V_{k+1}}{\partial \bb}\bigl(T(\cdots);\, \bc\bigr) \cdot \frac{\partial T(\bb, \bs_k^\star, \by_k;\, \bc)}{\partial \bb}}_{\text{(II) belief-update through }\bb}\, \biggr]
\label{eq:dV_db_unfolded}
\end{align}
with terminal condition
\begin{equation}
\frac{\partial V_K}{\partial \bb}(\bb;\, \bc) \;=\; \frac{\partial \Phi_0^{\mathrm{belief}}(\bb;\, \bc)}{\partial \bb}.
\label{eq:dV_db_terminal}
\end{equation}

Equation~\eqref{eq:dV_db_unfolded} is a proper recursion: $\partial V_k / \partial \bb$ at the current belief is expressed in terms of $\partial V_{k+1} / \partial \bb$ at child beliefs $T(\bb, \bs_k^\star, \by_k;\, \bc)$, initialized by Equation~\eqref{eq:dV_db_terminal}.  It runs on the same DP graph and the same backward sweep as the value recursion itself.

\paragraph{Structural comparison to $\partial V_k / \partial \bc$.}  Two observations:
\begin{itemize}[nosep, leftmargin=*]
\item \textbf{Two terms survive, not three.}  Equation~\eqref{eq:dV_unfolded} (the $\bc$-gradient) has three contributions because $\bc$ enters in three places: the outcome distribution, the belief update, and the recursive $V_{k+1}$.  Equation~\eqref{eq:dV_db_unfolded} (the $\bb$-gradient) has only two, because $\bb$ enters only in the outcome distribution (through marginalization) and in the belief update (through Bayes' rule).  There is no ``recursive-through-$\bb$'' term separate from the belief-update path: $V_{k+1}$ only receives $\bb$ via $T$, so the recursive dependence already lives inside term (II).
\item \textbf{Both recursions share the same DP graph.}  At every reachable $(k, \bb)$ where $V_k$ is tabulated, both $\partial V_k / \partial \bc$ (Equation~\eqref{eq:dV_unfolded}) and $\partial V_k / \partial \bb$ (Equation~\eqref{eq:dV_db_unfolded}) are evaluated as cotangents on the same graph.  In practice, reverse-mode AD returns both simultaneously as intermediate cotangents during the single backward sweep through the argmax-frozen recursion.
\end{itemize}

\paragraph{The Jacobian $\partial T / \partial \bb$.}  The belief-update Jacobian in term (II) is the Jacobian of Bayes' rule.  For discrete $\bx$ with belief vector $\bb = (b_1, \ldots, b_{K_\Phi})$, the update is $T(\bb)(x) = b(x)\, p(\by \mid x, \bs; \bc) / Z$ with $Z = \sum_{x'} b(x')\, p(\by \mid x', \bs; \bc)$, and a short calculation with the quotient rule gives
\begin{equation}
\frac{\partial T(\bb)(x)}{\partial b(x')} \;=\; \frac{p(\by \mid x, \bs; \bc)}{Z}\, \Bigl[\,\delta_{xx'} \;-\; \frac{b(x)\, p(\by \mid x', \bs; \bc)}{Z}\,\Bigr].
\label{eq:dT_db}
\end{equation}
One almost never writes Equation~\eqref{eq:dT_db} out in code; reverse-mode AD composes it automatically through the division in Bayes' rule.  The formula is useful for checking structure: $\partial T / \partial \bb$ is a rank-one perturbation of a diagonal matrix, which means the belief-update term (II) in Equation~\eqref{eq:dV_db_unfolded} has a simple form (a marginal-likelihood-weighted sum of a diagonal and a rank-one contribution).

\paragraph{When the recursion collapses.}  In the count-tuple formulation of §\ref{sec:qubit}, the ``belief'' is replaced by the discrete integer tuple $\mathbf{n}$, and the ``belief update'' is the combinatorial operation of appending $(n_\text{shot}, m)$ to bin $j$.  $\partial / \partial \mathbf{n}$ does not exist as a gradient (the tuple is discrete), and $\partial T / \partial \mathbf{n}$ is trivially null.  Equation~\eqref{eq:dV_db_unfolded} simply does not apply --- which is fine, because Equation~\eqref{eq:dV_unfolded}'s term (ii) also vanishes in that representation and all $\bc$-dependence flows through terms (i) and (iii) as already noted.  The $\partial V_k / \partial \bb$ recursion is the machinery that fills in term (ii) when the belief \emph{is} treated as a continuous-argument object, which is what the Case~C Gaussian-moment DP of the companion paper actually does.

\subsection{The operational recipe: freeze the argmax, then AD}

In practice one does \emph{not} implement Equation~\eqref{eq:dV_unfolded} by hand.  Instead the envelope theorem lets reverse-mode AD do the entire calculation in a single sweep, provided the argmax is frozen first.  The recipe:

\begin{enumerate}[nosep, leftmargin=*]
\item \textbf{Forward pass (no AD).}  Run the backward induction from $k = K$ down to $k = 0$, populating $V_k(\mathbf{n};\, \bc)$ in a memoization dictionary.  At each reachable $(k, \mathbf{n})$, \emph{record the argmax action $\bs_k^\star(\mathbf{n};\, \bc)$} alongside the value.  No AD tape is active during this pass.
\item \textbf{Gradient pass (AD-traced).}  Re-evaluate $V_0(p_0;\, \bc)$ by traversing the DP graph a second time --- but substitute the previously recorded $\bs_k^\star$ in place of the $\max_{\bs_k}$.  The expression being evaluated is now argmax-free:
\begin{equation}
V_0(p_0;\, \bc) \;=\; \sum_{\text{paths}} \Bigl[\prod_{k=1}^K \Pr(\by_k \mid \bb_{k-1}, \bs_k^\star, \bc)\Bigr]\, \Phi_0^{\mathrm{belief}}(\bb_K;\, \bc).
\label{eq:argmax_frozen}
\end{equation}
By Equation~\eqref{eq:envelope_tutorial} the value computed is numerically identical to the forward-pass value.
\item \textbf{Reverse-mode AD.}  Wrap Equation~\eqref{eq:argmax_frozen} in \texttt{Zygote.pullback} (or an equivalent reverse-mode AD system) and apply the pullback.  Because every primitive in the expression is either closed-form in $\bc$ or $\bc$-free, AD returns the exact $\partial V_0 / \partial \bc$ without any custom rule.
\end{enumerate}

The essential point: \emph{freezing the argmax turns the Bellman recursion into an ordinary differentiable expression}.  Without that step, AD would try to differentiate through the $\max$, producing subgradient terms that the envelope theorem already tells us are zero at the optimum.  With that step, the non-smooth $\max$ disappears from the AD graph entirely.

\subsection{What is \texorpdfstring{$\bc$}{c}-dependent in the qubit problem}

To make Equation~\eqref{eq:dV_unfolded} concrete: in the Ramsey flux sensor of §\ref{sec:qubit}, only one primitive carries $\bc$ in the entire DP --- the Ramsey single-shot likelihood
\begin{equation*}
p(\bx, \tau;\, \bc) \;=\; \tfrac{1}{2} + \tfrac{1}{2}\,\mathrm{th}(\bx, \bc)\, e^{-A(\bx, \bc)\,\tau - B(\bx, \bc)^2 \tau^2}\, \cos\bigl[\Delta\omega(\bx, \bc)\, \tau\bigr],
\end{equation*}
whose four rate/coefficient parameters $(\mathrm{th}, A, B, \Delta\omega)$ are closed-form functions of the seven hardware components of $\bc$.  Every other object in the DP inherits its $\bc$-dependence from this single primitive:

\begin{itemize}[nosep, leftmargin=*]
\item The marginal outcome probability $\Pr(m \mid \mathbf{n}, \tau, n_\text{shot};\, \bc)$ (Equation~\eqref{eq:marginal_outcome}) is a finite sum of Binomial factors evaluated at $p(\bx_i, \tau;\, \bc)$; its $\bc$-gradient comes through $p$.
\item The belief $\bb(\bx \mid \mathbf{n};\, \bc)$ (Equation~\eqref{eq:belief_from_counts}) is a normalized product of Ramsey factors; its $\bc$-gradient is the log-likelihood-ratio score weighted by the unnormalized posterior and renormalized.
\item The terminal reward $V_K(\mathbf{n};\, \bc) = -\mathrm{Var}(\bx \mid \mathbf{n};\, \bc)$ depends on $\bc$ only through its dependence on $\bb$.
\item The count-tuple state space, the action grid $\{(\tau_j, n_\text{shot}) : j, n_\text{shot}\}$, and the combinatorial child-tuple update $\mathbf{n} \mapsto \mathbf{n}'$ are $\bc$-free.
\end{itemize}

Consequently the AD graph for $\partial V_0 / \partial \bc$ contains a single $\bc$-dependent primitive (the Ramsey function), composed through the DP tree with the chain rule doing all of the bookkeeping.  No custom \texttt{rrule} for the Bellman $\max$ operator is needed --- that is the entire operational payoff of the envelope theorem.

\bigskip

The interested reader should turn to Sections~3 and~4 of the companion paper for the envelope-theorem statement in its full generality and the relaxation hierarchy that kicks in when the inner DP is not tractable even once.  This tutorial has covered only the tractable-inner-DP case, which corresponds to the paper's rungs~1 and~2 (Case~Studies~A and~B).

\section{Further reading and standard references}
\label{sec:reading}

\begin{itemize}[nosep, leftmargin=*]
\item \textbf{Classical DP.}  Bellman, \emph{Dynamic Programming}, 1957.  Bertsekas, \emph{Dynamic Programming and Optimal Control}, volumes I and II, especially the finite-horizon chapters.
\item \textbf{POMDPs.}  Kaelbling, Littman \& Cassandra, ``Planning and acting in partially observable stochastic domains,'' \emph{Artificial Intelligence}, 1998.  Smallwood \& Sondik, ``The optimal control of partially observable Markov processes over a finite horizon,'' \emph{Operations Research}, 1973.
\item \textbf{Bayesian optimal experimental design (one-step).}  Chaloner \& Verdinelli, ``Bayesian experimental design: a review,'' \emph{Statistical Science}, 1995.  Lindley, \emph{Making Decisions}, 1971.
\item \textbf{Sequential OED.}  Huan \& Marzouk, ``Sequential Bayesian optimal experimental design,'' \emph{Journal of Computational Physics}, 2016.  Kreucher \& Kastella, ``Sensor management using an active sensing approach,'' \emph{Signal Processing}, 2005.
\item \textbf{Reinforcement learning (the large-$K$, neural-policy regime).}  Sutton \& Barto, \emph{Reinforcement Learning: An Introduction}, 2018.
\end{itemize}

\appendix

\section{On \texorpdfstring{$\Phi_0$}{Phi\_0} and the reward conversion from pointwise to belief-level}
\label{app:phi0}

The pointwise reward $\Phi(\bx, \bs_{1:K};\, \bc)$ takes $\bx$ as an argument; the belief-level reward the DP actually consumes is $\E_{\bx \mid \bb_K}[\Phi_0(\bx, \bb_K;\, \bc)]$, which takes only $\bb_K$ and $\bc$.  The conversion is exactly one integration against the posterior.  For the two rewards used in this tutorial:

\begin{itemize}[nosep, leftmargin=*]
\item \textbf{Posterior variance.}  $\Phi(\bx, \bs_{1:K}; \bc) = -(\bx - \hat{x}(\bb_K))^2$ with $\hat{x}(\bb_K) = \E[\bx \mid \bb_K]$ the posterior mean.  Integrating gives $\E_{\bx \mid \bb_K}[-(\bx - \hat{x})^2] = -\mathrm{Var}(\bx \mid \bb_K)$, a pure functional of $\bb_K$.
\item \textbf{Mutual information.}  $\Phi_0(\bx, \bb_K; \bc) = \log \bb_K(\bx) - \log p_0(\bx)$.  Integrating gives $\E_{\bx \mid \bb_K}[\log(\bb_K/p_0)] = \mathrm{KL}(\bb_K \Vert p_0)$.  Averaging further over $\bb_K$ under the prior gives the mutual information $I(\bx; \by_{1:K})$.
\end{itemize}

A third case (Case~C of the companion paper) uses $\Phi_0 = \tfrac{1}{2} \log \det J_N(\bx, \bs_{1:K}; \bc)$, where the Fisher-information matrix $J_N$ depends on the action trajectory explicitly.  In that case the ``belief'' is augmented to carry the accumulated $J_N$ as an extra piece of state, so that the action-dependence folds into the belief and $\Phi_0$ becomes a function of the augmented belief alone.